\documentclass[a4paper, 11pt]{article}
\pdfoutput=1 % if your are submitting a pdflatex (i.e., if you have
\usepackage{jheppub} % for details on the use of the package, please
\usepackage{amsmath}
\usepackage{amssymb}
\usepackage{graphicx,xcolor}
\usepackage{subfig} 
\usepackage{empheq}
\usepackage{esint}  %\usepackage{pxfonts}
\usepackage{hyperref}
\usepackage[shortlabels]{enumitem}
\usepackage{tikz, pgf} 
\usetikzlibrary{arrows}
\usepackage{pgfplots}

\definecolor{pink}{HTML}{FDDBC7}
\definecolor{red}{HTML}{B2182B} 
\definecolor{blue}{HTML}{4393C3}

\newcommand{\beq}{\begin{equation}}
\newcommand{\eeq}{\end{equation}} 
\newcommand{\bea}{\begin{eqnarray}}
\newcommand{\eea}{\end{eqnarray}}

\newcommand{\fr}[2]{\frac{#1}{#2}}

\newcommand{\dl}{\delta}
\newcommand{\w}{\omega}

\newcommand{\be}{\beta}

\newcommand{\X}{X}
\newcommand{\qR}{q_{_R}}
\newcommand{\qL}{q_{_L}}

\newcommand{\qbP}{\bar{q}_{_P}}
\newcommand{\qbF}{\bar{q}_{_F}}

\newcommand{\qa}{q_{a}}
\newcommand{\qd}{q_{d}}

\begin{document}

\title{Nonlinear Langevin dynamics via holography}  
\author{Bidisha Chakrabarty,}
% \email{bidisha.chakrabarty@icts.res.in}
\author{Joydeep Chakravarty,}
% \email{joydeep.chakravarty@icts.res.in}
\author{Soumyadeep Chaudhuri,}
% \email{soumyadeep.chaudhuri@icts.res.in}
\author{Chandan Jana,}
% \email{chandan.jana@icts.res.in}
\author{R. Loganayagam,}
% \email{nayagam@icts.res.in}
\author{Akhil Sivakumar.}

\emailAdd{bidisha.chakrabarty@icts.res.in}
\emailAdd{joydeep.chakravarty@icts.res.in}
\emailAdd{chaudhurisoumyadeep@gmail.com} 
\emailAdd{chandan.jana@icts.res.in} 
\emailAdd{nayagam@icts.res.in}
\emailAdd{akhil.sivakumar@icts.res.in}

\affiliation{International Centre for Theoretical Sciences (ICTS-TIFR),
Tata Institute of Fundamental Research,
Shivakote, Hesaraghatta,
Bangalore 560089, India.}
\begin{abstract}
{In this work, we consider non-linear corrections to the Langevin effective theory of a heavy quark moving through a strongly coupled CFT plasma. In AdS/CFT, this system can be identified with that of a string stretched between the boundary and the horizon of an asymptotically AdS black brane solution. We compute the Feynman-Vernon influence phase for the heavy quark by evaluating the Nambu-Goto action on a doubled string configuration. This configuration is the linearised solution of the string motion in  the doubled black brane geometry which has been proposed  as the holographic dual of a thermal Schwinger-Keldysh contour of the CFT.  Our expression for the influence phase passes non-trivial consistency conditions arising from the underlying unitarity and thermality of the bath. The local effective theory obeys the recently proposed  non-linear fluctuation dissipation theorem
relating the non-Gaussianity of thermal noise to the thermal jitter in the damping constant. This furnishes a non-trivial check for the validity of these relations derived in the weak coupling regime.}
\end{abstract}
\maketitle 
\tableofcontents

\section{Introduction}
The Langevin theory, describing the motion of a particle coupled to a thermal bath, is the simplest example of an open quantum system.
The linear version of this theory with a linear damping $\gamma$ and Gaussian noise $\mathcal{N}^i$, described by the equation
\beq
\frac{d^2q^i}{dt^2}+\gamma \frac{dq^i}{dt} = \langle f^2\rangle \mathcal{N}^i \ ,
\eeq
serves as a textbook example of non-equilibrium statistical mechanics. Here $ \langle f^2\rangle$ is the variance
of the force per unit mass which controls the Gaussian ensemble for $\mathcal{N}^i$ :
\beq
\text{Probability}[\mathcal{N}^i] \sim \exp\left\{ - \frac{\langle f^2\rangle}{2}\int dt\ \mathcal{N}^i \mathcal{N}^i\right\} \ .
\eeq
The classic result of fluctuation dissipation theorem then asserts that this variance of the fluctuation $ \langle f^2\rangle$ is directly proportional
to the coefficient of linear drag/dissipation $\gamma$. From the point of view of open quantum systems, the emergence of this linear Langevin behaviour
has been well-explored beginning with the classic works of  Schwinger \cite{Schwinger:1960qe}, Feynman-Vernon \cite{Feynman:1963fq,Vernon:1959pla} and Caldeira-Leggett \cite{Caldeira:1982iu,Caldeira:1982uj}. It is now a 
paradigmatic example in the study of open quantum systems\cite{Breuer:2002pc,carmichael2009open,kamenev_2011}.

Despite the great theoretical and experimental successes of this model, it is clear that the picture of a linear dissipation with Gaussian noise can only be
an approximation to real systems. Any actual system exhibits nonlinearities in dissipation and non-Gaussianities in its noise. Thus, it is of great physical
interest to ask whether one could extend the linear Langevin theory to a \emph{non-linear} Langevin theory. Rather than adding ad hoc terms into the equation
above, one may ask, in the spirit of effective theory : what is the most general and universal non-linear extension which describes the dynamics of the Brownian particle?

As we review below, this question can be systematically addressed using the Schwinger-Keldysh formalism. The set of all terms that could be added to the 
theory of a Brownian oscillator, up to one derivative order  and till cubic order in amplitudes were classified in \cite{Chakrabarty:2018dov}. This has recently been extended to quartic 
dissipative oscillators by the authors of \cite{Chakrabarty:2019qcp}. The resultant non-linear Langevin theory with these additional terms describes the universal dynamics of 
a Brownian oscillator weakly coupled to a thermal bath. In fact, general sum rules could be written down which relate the parameters of the non-linear
Langevin theory and the thermal spectral functions/correlators  of the bath degrees of freedom \cite{Chakrabarty:2018dov}. Further, this model provides an arena where 
the consequences of microscopic time-reversal invariance and the non-linear generalisations of Onsager relations could be explored. New non-linear
generalisations of the fluctuation dissipation theorem emerge within this context relating the non-Gaussianities of the fluctuation ensemble to thermal jitter in the damping 
constant. This result puts in a systematic footing some of the previous attempts at generalising FDT (see \cite{efremov1969fluctuation,PhysRevA.18.2725,1981PhyA..106..443B,Wang:1998wg,Dubkov_2009}).

While the results of these works generalise Langevin theory and its characteristic features in interesting new directions, the validity of their analysis rest crucially
on treating the system-bath coupling perturbatively (Born approximation) and assuming that bath correlators decay exponentially fast resulting in a local effective
 theory for the system (Markov approximation). A typical example discussed in these works is that of a system oscillator non-linearly coupled to a set of bath oscillators
 in thermal equilibrium. The frequency spectrum/coupling of bath oscillators should then be chosen appropriately and the number of bath oscillators should be very large
 so as to guarantee fast decay of correlations. It is interesting to ask whether the structure of non-linear Langevin theory  holds true beyond  such weakly coupled systems in the
 `Born-Markov' regime. 
 
 In this note, we will answer this question in affirmative by studying a strongly coupled model in detail. The model we study is that of a heavy quark moving in  a bath of strongly coupled, thermal Yang-Mills plasma (or more generally thermal plasma of a strongly coupled conformal field theory (CFT)). The method we use to tackle this strongly coupled system is AdS/CFT duality, which maps the quark in the plasma problem to that of a string stretching from AdS boundary and probing a high temperature black brane \cite{deBoer:2008gu,Son:2009vu,Herzog:2006gh,CaronHuot:2011dr}. 
 
 The boundary endpoint of the string then represents the
heavy quark. The  string exhibits a random motion  due to the Hawking radiation of the transverse string modes.  These string modes also fall into the black brane thus resulting in  dissipation of energy. The combination of Hawking and  in-falling string modes then induce a Brownian motion for the heavy quark endpoint. We will use this system, widely studied within the holography literature \cite{deBoer:2008gu,Son:2009vu,Herzog:2006gh,Gubser:2006bz,Atmaja:2010uu,Herzog:2006se}, to compute the non-linear corrections to the quark dynamics.

The quark dynamics is corrected by terms generated in the Schwinger-Keldysh effective action while integrating out the bath degrees of freedom. Following Feynman-Vernon \cite{Feynman:1963fq,Caldeira:1982uj}, we will refer to these contributions to the effective action as the `influence phase'. The exponential of this influence phase can also be thought of as the real-time analog of a Wilson line in the gauge theory, albeit along the Schwinger-Keldysh time contour\footnote{We thank Amit Sever for pointing this out to us.}. In our case, the computation of the influence phase reduces in large N, strong coupling limit 
to  constructing a doubled string configuration that probes the holographic dual of the SK contour. The configuration we seek  is the linearised solution of the  
non-linear sigma model with a double black brane spacetime as the target space. This doubled black brane target space is  the  dual of the Schwinger-Keldysh contour in the CFT.

Given that we still lack a complete formalism which can address real time questions within string theory in general and holography in particular, some clarifying remarks are in order.
The doubled black brane spacetime that we use was  first proposed by van Rees \cite{vanRees:2009rw} (based on previous constructions of holographic thermo-field double by Son-Herzog \cite{Herzog:2002pc}, study by Son-Teaney\cite{Son:2009vu}
and especially the formalism elaborated by  Skenderis-van Rees \cite{Skenderis:2008dh,Skenderis:2008dg}). It was used to justify the ingoing prescription widely used in the literature \cite{Barnes:2010jp,Son:2002sd}. Crucially, the near-horizon structure
of this doubled  black brane spacetime was recently clarified by Glorioso-Crossley-Liu \cite{Glorioso:2018mmw}. \footnote{See \cite{deBoer:2018qqm} for a similar construction.}

Let us elaborate: such finite temperature  Schwinger-Keldysh saddle points have complicated near horizon structure  where the two solutions of SK contour meet, which is  a source of potential IR divergences, especially when non-linear corrections are taken into account \cite{Atmaja:2010uu}. In this context, a correct treatment of the outgoing  Hawking modes is essential to  derive unambiguous results within  the holographic Schwinger-Keldysh formalism.  This issue has long been an obstruction to constructing a full doubled configuration and extracting non-linear corrections unambiguously. As mentioned above, only recently has  the proper handling of the near horizon structure  in holographic Schwinger-Keldysh  saddle points been clarified by Glorioso-Crossley-Liu \cite{Glorioso:2018mmw}. We implement their near-horizon prescription  to fix the string saddle point. 
 
 The string configuration we construct is a simple extension of the 
 Son-Teany configuration \cite{Son:2009vu} and we demonstrate that it cures the IR divergences, at least at the leading order. Our computation
can hence be thought of as a non-trivial check of their prescription. Our work also  extends the holographic Schwinger-Keldysh formalism to interacting fields for the first time to show
that it gives sensible results.

The essential ideas behind the Glorioso-Crossley-Liu prescription that are relevant to this work are two in number : first, one works within the ingoing Eddington-Finkelstein chart in which the ingoing modes are
analytic whereas the outgoing Hawking modes show up as non-analytic solutions. Second,   two copies of the black brane solution are taken and  the region interior to their stretched horizons are removed. The two geometries are then stitched together by Wick rotating the tortoise co-ordinate. The coefficient of the Hawking modes are then fixed by interpolating between the two branches through this intermediate geometry. We show that this prescription when implemented at the level of world sheet, gives finite results consistent with CFT expectations.

After that technical aside on holographic SK contours, let us return to the description of the physical system under study. The origin of the non-linear corrections to the quark  influence phase lies in   the non-linearities of the world-sheet sigma model. These corrections encode the physics of 
world-sheet interactions among the string modes in the black brane background. These world-sheet processes are, via AdS/CFT, dual to the strong interactions between 
the heavy quark and the surrounding CFT plasma which, on the CFT side, results in non-linearities in the Brownian motion of the quark. In particular, the interacting string modes 
lead to  non-Gaussianity in the Hawking noise  felt by the string as well as a jitter in the rate of energy in-fall into the black brane. Holography maps these physical processes to 
fluctuation/dissipation in the  non-linear Langevin dynamics of the heavy quark.
 
On examining the influence phase computed from  the string configuration, we find that the universal non-linear
 Langevin dynamics derived from weakly coupled intuition, continues to describe the non-linearities in this strongly coupled system.
  The holographic Brownian motion that we study serves as a good model to illustrate the emergence of the new non-linear  fluctuation dissipation theorem  controlling the 
 non-Gaussianities of the noise. Further, the simplicity of the model allows us to compute the full influence phase at the quartic level which capture the nonlinear effects
 of the bath on the particle beyond derivative expansion/Markovian approximation.  
 
 In addition, we examine how the thermality of the bath controls the structure of the non-linear influence phase. Thermality constrains the bath correlators
via Kubo-Martin-Schwinger (KMS) conditions which impose appropriate periodicity in imaginary time.  The KMS structure of the correlators
along with unitarity of bath   in turn  constrain the structure of the non-linear influence phase felt by the heavy quark \cite{Henning:1993gh,Chu:1993nc,Henning:1995sm,
Hou:1998yc,Wang:1998wg,Chaudhuri:2018ymp}. This
structure becomes manifest in the Retarded-Advanced(RA) basis \cite{Aurenche:1991hi,vanEijck:1992mq,Baier:1993yh,vanEijck:1994rw,Chaudhuri:2018ymp} which naturally emerges out of our holographic construction.
  
Apart from being a simple model where one can study non-linear fluctuation dissipation theorems, the Brownian string problem is also one of 
the simplest  non-equilibrium set up within string theory. We hope that our study would eventually lead to a practical and systematic formalism to 
address real time questions within string theory. We postpone further remarks about future directions to the discussion section.

We will conclude this introduction by summarising what we do in the subsequent sections. We begin by reviewing the non-linear Langevin theory 
of the quartic oscillator in \S\ref{sec:quartic} with a focus on its description in terms of SK effective action. Next section \S\ref{sec:holoSKB} 
starts with a  review of the black brane background and its holographic SK contour counterpart. This is followed by a description of the main system of study, viz., the Brownian string.
We again start with a single copy description and then lift the results to the holographic SK contour. In the same section, we also describe how the linearised solution can be used to compute the non-linear corrections to the influence phase. This is followed in section \S\ref{sec:nonlinLang} by a derivation of the non-linear influence phase  along with its derivative expansion. 
We conclude with a discussion on future directions in \S\ref{sec:discussion}.

We have a series of appendices summarising the technical details of our calculations. Appendix \ref{app:DerivExp} expounds on the structure of the 
string solution to arbitrary orders in derivative expansion. This is followed by appendix \ref{app:AdS3} where we consider a particular example of string 
probing BTZ black brane in AdS$_3$. Exact expressions for influence phase to all orders in derivative expansion can be written down in this case.

\section{Non-linear Langevin theory}\label{sec:quartic} 
Our goal in this section is to  introduce the non-linear Langevin theory for the Brownian particle and its description within Schwinger-Keldysh formalism.
The discussion here is based on the results of \cite{Chakrabarty:2018dov,Chakrabarty:2019qcp} which we will refer the reader for a more detailed 
description of the models under consideration. The reader will find in those works a detailed justification of how the non-linear Langevin effective theory arises generically out of weakly coupled baths along with examples drawn from  simple bath models. 

Consider a Brownian particle moving in a spacetime 
of dimension $d$. We denote by $q^i$ the spatial position of this particle, where $i=1,2,\ldots (d-1)$ denote spatial indices. In our context, we are interested in a heavy quark coupled to a CFT$_d$ plasma at a temperature $T$. This quark then feels a variety of forces, which at a
macroscopic effective theory level result in energy/momentum dissipation and fluctuations. We will describe this by a stochastic ordinary differential equation 
for the position of the quark :
\begin{equation}\label{eqn:StochasticEOM}
\begin{split}
\mathcal{E}^i[q,\mathcal{N}]\equiv &\ \frac{d^2q^i}{dt^2}+\Big(\gamma+\zeta_{\gamma}\mathcal{N}^2\Big) \frac{dq^i}{dt} -\langle f^2\rangle \mathcal{N}^i+\ldots =0.
\end{split}
\end{equation}
Here $\mathcal{N}^i$  is a thermal Non-Gaussian noise drawn from a  probability distribution $\widetilde{P}[\mathcal{N}]$ invariant under rotations and parity. The form of this distribution will be specified in the next subsection. The quantity $\mathcal{N}^2$ is the rotationally invariant quadratic combination of the noise fields defined below:
\begin{equation}
\begin{split}
\mathcal{N}^2\equiv \mathcal{N}^i \mathcal{N}^i.
\end{split}
\end{equation}
We use $\langle f^2\rangle$ to denote the strength of the additive noise in the equation of motion, and $\gamma$ to denote the damping coefficient resulting in the drag force. The coefficient $\zeta_{\gamma}$ is the strength of a thermal jitter in the damping coefficient. 

We immediately note that the system under consideration has a variety of symmetries including translational, rotational and reflection invariance. These symmetries drastically reduce the number of non-linear corrections to the linear theory that can appear in general. This is in  contrast to \cite{Chakrabarty:2018dov,Chakrabarty:2019qcp}, where a non-linear Brownian \emph{oscillator} without translational invariance was studied. 
%As we will argue below, the above equation then represents the most general non-linear Langevin equation to quartic order in $q$ and upto first order in time derivatives.

In the following subsection, the non-linear Langevin dynamics described by Eqn.\eqref{eqn:StochasticEOM} will be shown to be dual to a Schwinger-Keldysh effective theory. Based on this duality, we will argue that the $\zeta_\gamma$-term in this equation represents 
the universal leading correction (consistent with the symmetries) to the standard linear Langevin dynamics.

\subsection{Stochastic path integral}
We would like to now write down an effective action/path integral  which computes correlators in this stochastic theory. This can be done by using the technique of Martin-Siggia-Rose \cite{1973PhRvA...8..423M}-de Dominicis-Peliti \cite{1978PhRvB..18..353D}-Janssen \cite{1976ZPhyB..23..377J}. We will do this via the following steps. In the process, we will also specify the form of the probability distribution of the thermal noise.

In the first step, we  write down  a path integral (see, for example, Sec.(4.3) of \cite{kamenev_2011} and Sec.(4.6) of \cite{ZinnJustin:1989mi})
\beq
\begin{split}
\mathcal{Z}=&\int [D\mathcal{N}] [D q_a]\ \det\Bigg[\frac{\delta\mathcal{E}^i[q_a(t),\mathcal{N}(t)]}{\delta q_a^j(t^\prime)}\Bigg]\widetilde{P}[\mathcal{N}] \prod_i\delta(\mathcal{E}^i[q_a,\mathcal{N}])\\
=&\int [D\mathcal{N}] [D q_a]\ \det\Bigg[\delta_{ij}\Big\{\frac{\partial^2}{\partial t^2}+\Big(\gamma+\zeta_{\gamma}\mathcal{N}^2\Big)\frac{\partial}{\partial t}\Big\}\delta(t-t^\prime)\Bigg]\widetilde{P}[\mathcal{N}] \prod_i\delta(\mathcal{E}^i[q_a,\mathcal{N}])\ .\\ 
\end{split}
\eeq
Here $\delta(\mathcal{E}^i[q_a,\mathcal{N}])$ indicates that we choose to average over all solutions of our stochastic ODE, with an appropriate weight for the noise field. This path integral
can then be used to compute noise-averaged correlators of the non-linear Langevin system.  

The determinant in such stochastic path integrals are usually  dealt with via a retarded/Ito regularisation or via ghosts \cite{kamenev_2011, ZinnJustin:1989mi, 2011arXiv1102.1581H, 2010arXiv1009.5966C}.\footnote{We thank the referee for drawing our attention to this point.} In our problem, given that the determinant is purely a functional of the noise fields $\mathcal{N}^i$, the determinant serves to merely renormalise the noise distribution. We define this renormalised probability distribution  to be 
\beq
\begin{split}
P[\mathcal{N}]&\equiv\det\Bigg[\delta_{ij}\Big\{\frac{\partial^2}{\partial t^2}+\Big(\gamma+\zeta_{\gamma}\mathcal{N}^2\Big)\frac{\partial}{\partial t}\Big\}\delta(t-t^\prime)\Bigg]\widetilde{P}[\mathcal{N}] .
\end{split}
\eeq
We assume that the form of this renormalised distribution is
\begin{equation}\label{eqn:NoisePDF}
P[\mathcal{N}]\ \propto\ \exp\Big[-\int dt\Big(\frac{\langle f^2\rangle}{2}\mathcal{N}^2+\frac{Z_I}{2}\left(\frac{d\mathcal{N}}{dt}\right)^2+\frac{\zeta_N}{4!}(\mathcal{N}^2)^2+\ldots\Big)\Big]\ ,
\end{equation}
where $\langle f^2\rangle$ denotes the variance of the distribution \footnote{Note that $\langle f^2\rangle$ also enters in the equation of motion as the coefficient of the additive noise.}, $\zeta_N$ parametrises its non-Gaussianity, and $Z_I$ characterises  its frequency dependence. This non-Gaussian distribution enters in the stochastic path integrals as shown below:
\beq
\begin{split}
\mathcal{Z}=&\int [D\mathcal{N}] [D q_a]P[\mathcal{N}] \prod_i\delta(\mathcal{E}^i[q_a,\mathcal{N}])\ .\\ 
\end{split}
\eeq

The formal expression above, as is usual with  path integrals,  should be thought of as a limit of a UV regulated expression in which appropriate counter-terms are added to cancel UV divergences. A common way to regulate the expression is to discretise the time steps, add counter-terms and take the time-step $\delta t \to 0$ limit. For example, to leading order in 
the non-linear couplings, the following counter-terms are sufficient to guarantee finite answers :
\beq
\begin{split}
[D\mathcal{N}] [D q_a] = \lim_{\delta t \to 0} [D\mathcal{N}]_{\delta t} [D q_a]_{\delta t} \exp\Big[\frac{1}{\langle f^2\rangle\delta t}\int dt\Big(\zeta_\gamma\mathcal{N}^i\frac{dq_a^i}{dt}+\frac{1}{4}\zeta_N\mathcal{N}^2\Big)\Big]\ .
\end{split}
\eeq

The second step in the MSR-DPJ method is  to introduce  an auxiliary field for each direction to deal with the functional delta function. This gives
\beq
\begin{split}
\mathcal{Z}=&\int [D\mathcal{N}] [D \qa][D \qd]\exp\Big[-i\ m_p \int dt\ \qd^i\mathcal{E}^i[\qa,\mathcal{N}]\ \Big]P[\mathcal{N}]\ . 
\end{split}
\eeq
Here $m_p$ denotes the mass of the Brownian particle which is introduced for dimensional reasons and $\qd^i$ is the auxiliary field enforcing the
stochastic equation of motion. In the third step, we perform the integral over the noise to get an effective action for the non-linear Langevin theory of the form
\beq
\begin{split}
\mathcal{Z}=&\int  [D \qa][D \qd]e^{i\int dt L_{SK}} \ , 
\end{split}
\eeq
where the effective Lagrangian takes the form
\beq\label{eqn:SKEffectiveAction}
\boxed{\begin{split}
L_{SK}\equiv \quad & m_p \frac{d\qa^i}{dt} \frac{d\qd^i}{dt} -m_p\gamma\frac{d\qa^i}{dt}\qd^i+m_p^3\zeta_\gamma\qd^2\frac{d\qa^i}{dt}\qd^i\\
&+i \left\{m_p^2\frac{\langle f^2\rangle}{2!}\qd^2 - m_p^2\frac{ Z_I}{2!}\left(\frac{d\qd}{dt} \right)^2+m_p^4\frac{\zeta_N}{4!}(\qd^2)^2\right\} +\ldots\ .
\end{split}}
\eeq
We have denoted the  effective Lagrangian by the symbol $L_{SK}$, since this Lagrangian can alternately be thought of as coming from solving the underlying quantum non-equilibrium
dynamics of the heavy quark moving in a CFT plasma using the Schwinger-Keldysh formalism, as we describe below.

\subsection{Heavy quark in SK formalism}
\begin{figure}[h!]
\centering
\begin{center}
\caption{Schwinger-Keldysh contour}
\scalebox{0.7}{\begin{tikzpicture}[scale = 1.5]
\draw[thick,color=violet,->] (-3,0 cm) -- (3,0 cm) node[midway,above] {\scriptsize{R}} ;
\draw[thick,color=violet,->] (3, 0 cm - 0.5 cm) -- (-3, 0 cm - 0.5 cm)node[midway,above] {\scriptsize{L}}  ;
\draw[thick,color=violet,->] (3, 0 cm) arc (90:-90:0.25) ;
\end{tikzpicture}}
 \label{fig:SK contour}
 \end{center}
\end{figure}
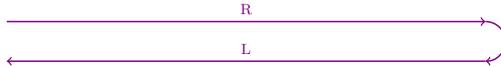

A dual perspective on Brownian motion and Langevin theory is obtained by thinking about their quantum microscopic origins. The microscopic description of the heavy quark is 
in terms of a reduced density matrix which evolves non-unitarily due to the interaction with the CFT plasma bath. To keep track of the density matrix and its evolution, one needs
to move to the Schwinger-Keldysh(SK) formalism \cite{Schwinger:1960qe,Keldysh:1964ud,CHOU19851,Haehl:2016pec,Feynman:1963fq}.  One introduces in the SK formalism, two fields per degree of freedom to keep track of the evolution on the ket and the bra side of the density matrix. We define $\qR$ and $\qL$ as the right-moving/Ket and left-moving/Bra degrees of freedom in the Schwinger-Keldysh contour shown in Fig.\ref{fig:SK contour}.

The effective SK action for the heavy quark is obtained by starting with the SK path integral of the whole system and then systematically integrating out the CFT bath degrees of freedom.
The process of integrating out the bath introduces  corrections to the particle SK action. These corrections encode the effect of the particle-bath interaction in the macroscopic effective theory of the heavy quark. Following Feynman-Vernon \cite{Feynman:1963fq}, these corrections are referred to as the ``influence phase" of the heavy quark. In general, the influence phase of the quark
is determined by the connected parts (cumulants) of contour ordered correlators of the CFT operators that couple to the quark. In a strongly coupled CFT, a direct computation
of the influence phase on the CFT side is a near impossible task. We circumvent this difficulty by using AdS/CFT and working on the gravity side instead.

Even if we cannot compute the influence phase on the CFT side, one could deduce, based on general considerations, various structural properties of the effective action. The unitarity
and thermality of the CFT strongly constrain the  influence phase. Later, we will see how the answer from gravity manages to satisfy these constraints. 

On a related note, the local approximation of the influence phase also satisfies a set of constraints. While the influence phase due to the CFT bath is in general non-local, if the bath cumulants  decay sufficiently fast, one can viably expand the  influence phase in a derivative expansion. Even if the influence phase from the CFT is hard to compute, its local approximation
can be characterised using effective theory. We can write down the most general local expression consistent with microscopic constraints and parameterise our ignorance in terms of a set of phenomenological parameters.

The relevant microscopic constraints are three in number \cite{CHOU19851,kamenev_2011,Sieberer:2015svu,Avinash:2017asn} : 
\begin{enumerate}
\item The microscopic unitarity in the dynamics of the (particle+bath) combined system which in particular implies that the action should be zero if $\qR^i=\qL^i$ . This condition is necessary for the heavy quark's correlators to satisfy the largest time equation.
%This is called the Lindblad condition and it is a consequence of the largest time equation.
This condition is also responsible for the Lindblad structure of the master equation obeyed by  the heavy quark's density matrix \cite{Avinash:2017asn, Sieberer:2015svu}. 
\item The reality conditions which involve $\qR^i\leftrightarrow \qL^i$ exchange and
\item Invariance of the microscopic dynamics under constant translations, rotations and reflection of the $q^i$'s.
\end{enumerate}
If we define \cite{Keldysh:1964ud,CHOU19851,kamenev_2011}
\beq
\qa^i\equiv\frac{1}{2}(\qR^i+\qL^i),\ \qd^i\equiv(\qR^i-\qL^i) ,
\eeq
 the most general quartic Lagrangian (up to terms with single time derivatives) consistent with these constraints is given by $L_{SK}$ written down in the 
 previous subsection. Thus by completely different arguments than before, we arrive at the same result for the effective action of the Brownian particle. This leads us to  conclude that the nonlinear $\zeta_\gamma$-term in Eqn.\eqref{eqn:StochasticEOM} is 
the universal leading correction (consistent with the symmetries) to the standard linear Langevin dynamics.

 The mapping between the stochastic variables $\qa$ and $\qd$ vs $\qR$ and $\qL$ of the Schwinger-Keldysh description given by the expression above
establishes a duality between two different ways of viewing the Langevin theory : one as a stochastic theory and another as a macroscopic SK effective
theory for an open quantum system. To these two perspectives, gravity adds a third perspective in the form of an open string probing a black brane.

\subsection{Thermality, time reversal and fluctuation-dissipation relations}
In the previous subsection, we mentioned that the effects of the particle-bath interaction on the quark's dynamics are encoded in its influence phase. Such influence phases have been studied in detail for several microscopic models where a Brownian particle is weakly coupled to a thermal bath \cite{Caldeira:1982iu, PhysRevD.45.2843, PhysRevD.47.1576, Chakrabarty:2018dov, Chakrabarty:2019qcp}. In such scenarios, one can perturbatively expand the influence phase as a power series in the particle-bath coupling. The $n^{\text{th}}$ degree term in this expansion receives contributions from the n-point cumulants of the bath operator that couples to the particle \cite{Feynman:1963fq,Breuer:2002pc}. This relationship between the particle's influence phase and the bath's cumulants allows one to study the constraints imposed on the particle's effective dynamics due to microscopic symmetries of the bath.

An interesting example of such a symmetry is the time reversal invariance of the bath's microscopic dynamics. Such a symmetry in the bath relates the bath's correlators to their time-reversed counterparts. These relations between the bath's correlators, in turn, lead to relations between the coefficient functions of different terms in the non-local influence phase. In the derivative expansion of the influence phase, these relations reduce to certain reciprocal relations between the effective couplings. Such reciprocal relations were first explored by Onsager \cite{PhysRev.37.405, PhysRev.38.2265} and were later extended by Casimir \cite{RevModPhys.17.343}  for the quadratic couplings in the effective dynamics of a system with multiple degrees of freedom. The discovery of these  relations laid the foundations for understanding how the dynamics of an open system, which manifestly violates reversibility due to dissipation, is nevertheless constrained by microscopic time-reversal invariance in the environment.

Recently,   the Onsager-Casimir reciprocal relations were generalised to the cubic and  quartic couplings of a Brownian particle in \cite{Chakrabarty:2018dov} and \cite{Chakrabarty:2019qcp}. These papers demonstrated that under time reversal certain  Schwinger-Keldysh correlators of the bath get mapped to its Out of Time Order Correlators (OTOCs) \footnote{Such OTO correlators have been found to be useful for diagnosing chaos, thermalisation,many-body localisation, etc. in quantum systems. We refer the reader to  \cite{Swingle:2017jlh} for discussions on the importance of such OTOCs.}. As argued in these works, a convenient way to study the effects of such relations between the bath's correlators on the particle's dynamics is to extend its effective theory by including the contributions of the bath's OTOCs  \cite{Chaudhuri:2018ihk, Chakrabarty:2019qcp}. Such an extension of the effective theory requires the introduction of some new OTO couplings \cite{Chaudhuri:2018ihk, Chakrabarty:2019qcp} which encode  information about the bath's OTOCs. The relations between these OTOCs and Schwinger-Keldysh correlators due to microscopic reversibility in the bath constrains the particle's effective theory by connecting some of the additional OTO couplings to couplings appearing in the Schwinger-Keldysh effective theory of the particle \cite{Chakrabarty:2018dov, Chakrabarty:2019qcp}. 

Apart from the bath's microscopic reversibility, there is another source of relations between the particle's effective couplings, viz. the thermality of the bath. The fact that the bath is in a thermal state implies the existence of certain relations between its correlators which were first discussed by Kubo \cite{Kubo:1957mj}, Martin and Schwinger \cite{Martin:1959jp}. These relations (now commonly known as the KMS relations) connect thermal correlators which can be obtained from each other by cyclic permutation of insertions as shown below:
\beq
\langle O_1(t_1)O_2(t_2)\cdots O_n(t_n)\rangle=\langle O_n(t_n-i\beta)O_1(t_1)O_2(t_2)\cdots O_{n-1}(t_{n-1}) \rangle,
\eeq
where $O_1, O_2, \cdots O_n$ are $n$ operators of the bath, and $\beta$ is the inverse of the bath's  temperature (in units where the Boltzmann constant $k_B=1$). In the high temperature limit, such KMS relations between 2-point correlators of the bath lead to the following relation between two quadratic couplings in the particle's Schwinger-Keldysh effective action (see \eqref{eqn:SKEffectiveAction}):
\beq
\langle f^2\rangle=\frac{2}{\beta m_p}\gamma.
\eeq
From the point of view of the dual Langevin dynamics given in \eqref{eqn:StochasticEOM}, this implies the well-known fluctuation-dissipation relation (FDR) \cite{PhysRev.32.97, PhysRev.32.110, PhysRev.83.34} which connects the strength of the thermal random force acting on the particle to its damping coefficient. 

The authors of \cite{Chakrabarty:2018dov} and \cite{Chakrabarty:2019qcp} generalised this FDR  to  cubic and quartic OTO couplings of the particle respectively. This was achieved by examining the KMS relations that connect Schwinger-Keldysh correlators of the bath to its OTOCs by analytic continuation \cite{Haehl:2017eob}. KMS relations along with the generalised Onsager-Casimir relations implied a generalised FDR of the form : 
\beq\boxed{
\zeta_N=-\frac{12}{\beta m_p}\zeta_\gamma}\ .
\eeq
This FDR relates the non-Gaussianity in the thermal noise to the thermal jitter in the particle's damping coefficient.
A higher dimensional version of this relation applies to the stochastic dynamics introduced in \eqref{eqn:StochasticEOM} and \eqref{eqn:NoisePDF}, as we shall show via holography.

\section{Holographic Brownian motion}\label{sec:holoSKB}
We will now move on to our main system of interest : an open string hanging from the AdS boundary probing a black brane geometry. In particular, we want to 
compute the Schwinger-Keldysh path integral, which means that the black brane background should be doubled in some appropriate way. We will begin by describing
the geometry of  this doubled spacetime constructed according to the prescriptions of \cite{Glorioso:2018mmw,Skenderis:2008dh,Skenderis:2008dg,vanRees:2009rw}. We will later see how this target space yields sensible answers  for  the 
influence phase of the heavy quark.

\subsection{The black brane and the holographic SK contour}
Following \cite{Glorioso:2018mmw},
we begin by considering the asymptotically AdS$_{d+1}$ black brane, written in ingoing Eddington-Finkelstein co-ordinates 
\beq
ds^2= 2\ dv\ dr -r^2\left(1-\frac{r_h^d}{r^d}\right) dv^2+ r^2 dx^2_{d-1}\ .
\eeq
Here $r_h$ is the horizon radius. We have set the AdS radius to unity. The inverse Hawking temperature of this black brane is given by 
\beq
\beta\equiv\frac{4\pi}{dr_h} .
\eeq
We are interested in a string probing a doubled avatar of this spacetime, whose mathematical description is in terms of an appropriate
non-linear sigma model living on the string worldsheet. For this reason, we will be interested in re-writing this spacetime with other
more convenient co-ordinates.

It is often convenient to scale out the horizon radius by introducing a scaled radial co-ordinate $\rho\equiv \frac{r}{r_h}$. The above metric then becomes
\beq
ds^2= r_h^2\left\{ \frac{2}{r_h}\ dv\ d\rho - \rho^2\left(1-\frac{1}{\rho^d}\right) dv^2+ \rho^2 dx^2_{d-1}\right\}\ .
\eeq
The horizon in this co-ordinates is at $\rho=1$ and the AdS boundary is at $\rho\to \infty$. From now on, we will regulate this boundary to be at a cutoff surface given by $\rho=\rho_c$.

 For the purposes of studying the string non-linear sigma model, it is useful to shift further to a \emph{negative imaginary} radial co-ordinate $\xi$ defined in the exterior of the black brane via 
\beq
\frac{d\xi}{dr}  \equiv  \eta
    \fr{r^{d-4}}{r^d -r_h^d}\quad \text{or}\quad \frac{d\xi}{d\rho}  \equiv  \fr{d}{2 \pi i}
    \fr{\rho^{d-4}}{\rho^d -1}\ ,
\eeq
where $\eta$ is a negative imaginary constant defined by 
\bea
\eta &\equiv &r_h^3 \frac{d}{2\pi i }\ .
\eea
The co-ordinate $\xi$ shares many features of the tortoise co-ordinate and plays a similar role in the worldsheet theory. 
%The metric in the $\xi$ co-ordinate takes the form
%\beq
%ds^2= \left(\frac{d\xi}{dr}\right)^{-1}\left\{ 2\ dv\ d\xi -\eta \frac{dv^2}{r^2}\right\}+r^2 dx^2_{d-1}\ .
%\eeq
Note that at this juncture, we have not changed in anyway the geometry of the  black brane spacetime. For now, we are merely using an imaginary co-ordinate
to label points on a real Lorentzian spacetime.

In the next step, we would now like to double this black brane spacetime so as to construct a solution with CFT SK contour as its boundary \cite{Skenderis:2008dh,Skenderis:2008dg,vanRees:2009rw,Glorioso:2018mmw}. Such a spacetime will
then serve as the holographic SK contour which computes SK correlators. The geometry of this spacetime contains various non-trivial features both along the time
direction and the radial direction.

\begin{figure}
\begin{center}
\begin{tikzpicture}[rotate=90]

\draw[ultra thick,blue] (0,0) -- (5,0);
\draw[ultra thick,blue] (5,0) -- (5,-0.5);
\draw[ultra thick,blue] (5,-0.5) -- (0,-0.5);
%\draw[ultra thick,blue] (0,-0.5) -- (0,-2);

\node at (5.5,-0.2) {$v\to\infty$};
\node at (-0.2,-0.2) {$v=-\infty$};
\node at (2.5,0.5) {$M_R$};
\node at (2.5,-1) {$M_L$};
%\node at (0.5,-1.5) {$M_E$};

\end{tikzpicture}
\end{center}
\caption{\label{fig:fixedRho} Picture of the bulk at a fixed $\rho$ : The 2 vertical lines represent the constant $\rho$ sections of $M_R$ and $M_L$. These sections meet at the future-turning point ($v\rightarrow\infty$).}
\end{figure}
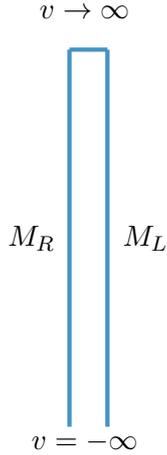

Let us begin by commenting on the geometry along time direction. The AdS boundaries of the holographic SK contour at $(\rho=\rho_c\pm i\varepsilon)$ asymptote to the
SK contour seen by the CFT. This boundary has a doubled  time contour  as illustrated in the fig.\ref{fig:fixedRho}. This structure is replicated at every fixed $\rho$ radial slice in the  bulk. 

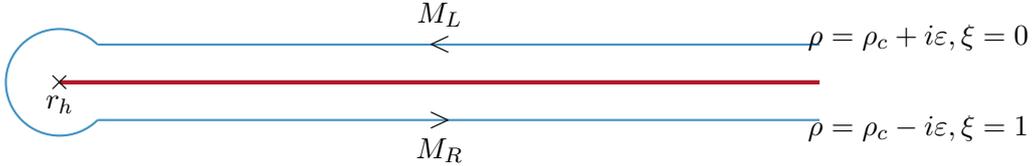
\begin{figure}
\begin{center}
\begin{tikzpicture}

\draw[red,ultra thick] (0,0) -- (10,0);
\node at (0,0) {$\times$};

\draw[blue, thick] (10,0.5) -- (0.5,0.5);
\draw[blue, thick] (0.5,0.5) arc (45:315:0.707);
\draw[blue, thick] (10,-0.5) -- (0.5,-0.5);

\node at (5,0.5) {$<$};
\node at (5,0.9) {$M_L$};
\node at (5,-0.5) {$>$};
\node at (5,-0.9) {$M_R$};
\node at (0,-0.3) {$r_h$};

\node at (11.3,0.6) {$\rho=\rho_c+i\varepsilon,\xi=0$};
\node at (11.3,-0.6) {$\rho=\rho_c-i\varepsilon,\xi=1$};

\end{tikzpicture}
\end{center}
\caption{\label{fig:fixedv} Picture of the bulk at a fixed $v$ : The red line is the branch cut, which starts at the horizon, $r_h$ and ends at the AdS boundary. The direction of the contour is from the upper segment to the lower one.} 
\end{figure}

The geometry along the radial direction is illustrated  in the fig.\ref{fig:fixedv}.  On the radial direction, the right and the left copies of the black brane (denoted by $M_R$ and $M_L$)
are stitched together at their stretched horizon (i.e, at $r<r_h$) by a `horizon cap'. This horizon cap region regulates the outgoing Hawking modes which blow up at the horizon in the ingoing EF co-ordinates. 

\begin{figure}
\begin{center}
\begin{tikzpicture}

\begin{scope}[shift={(0,0)},scale=0.8]
\draw (-2,-2) -- (-2,2) (2,-2) -- (2,2);

\draw[decoration = {zigzag,segment length = 1mm, amplitude = 0.3mm}, decorate,shift={(0,2)}] (-2,0) .. controls (0,-0.5) .. (2,0);
\draw[decoration = {zigzag,segment length = 1mm, amplitude = 0.3mm}, decorate,shift={(0,-2)}] (-2,0) .. controls (0,0.5) .. (2,0);
\draw[color=blue!80, fill=green!40, thick] (2,1.8) -- (0.2,0) -- (2,-1.8);
\draw[dashed,thick] (2,-1.2) -- ({0.25+0.3*cos(45)},0.3);
\draw[densely dashed,thick] ({0.25+0.3*cos(45)},0.3) arc (-40:-310:0.13);

\draw[dashed,thick] (2,-0.6) -- ({0.25+0.3*cos(45)+0.6*cos(45)*cos(45)},{0.3+0.6*sin(45)*sin(45)});
\draw[densely dashed,thick] ({0.25+0.3*cos(45)+0.6*cos(45)*sin(45)},{0.3+0.6*sin(45)*sin(45)}) arc (-40:-310:0.13);

\draw[dashed,thick] (2,0) -- ({0.25+0.3*cos(45)+1.2*cos(45)*cos(45)},{0.3+1.2*sin(45)*sin(45)});
\draw[densely dashed,thick] ({0.25+0.3*cos(45)+1.2*cos(45)*sin(45)},{0.3+1.2*sin(45)*sin(45)}) arc (-40:-310:0.13);

\draw (-2,-2) -- (2,2);
\draw (-2,2) -- (2,-2);
\node[rotate=-45] at ({0.15+1.2*cos(135)},{0.15+1.2*sin(135)}) {{\scriptsize{$\leftarrow V=0 \rightarrow$}}};
\node[rotate=45] at ({-0.15+1.2*cos(225)},{0.15+1.2*sin(225)}) {{\scriptsize{$\leftarrow U=0\rightarrow $}}};

\node at (-2.5,-2.3) {{\scriptsize{$ -\frac{\pi}{2}=V$}}};
\node at (2.5,2.3) {{\scriptsize{$ V=\frac{\pi}{2} $}}};

\node at (2.5,-2.3) {{\scriptsize{$U = -\frac{\pi}{2}$}}};
\node at (-2.5,2.3) {{\scriptsize{$ \frac{\pi}{2}=U $}}};

\node at (2.6,0) {{\scriptsize{$M_R$}}$\uparrow$};

\node at (0,-3) {$(a)$};
\end{scope}

\begin{scope}[shift={(5,0)},scale=0.8]

\draw (-2,-2) -- (-2,2) (2,-2) -- (2,2);

\draw[decoration = {zigzag,segment length = 1mm, amplitude = 0.3mm}, decorate,shift={(0,2)}] (-2,0) .. controls (0,-0.5) .. (2,0);
\draw[decoration = {zigzag,segment length = 1mm, amplitude = 0.3mm}, decorate,shift={(0,-2)}] (-2,0) .. controls (0,0.5) .. (2,0);
\draw[color=blue!80, fill=blue!40, thick] (2,1.8) -- (0.2,0) -- (2,-1.8);

\draw[dashed,thick] (2,-1.2) -- ({0.25+0.3*cos(45)},0.3);
\draw[densely dashed,thick] ({0.25+0.3*cos(45)},0.3) arc (-40:240:0.13);

\draw[dashed,thick] (2,-0.6) -- ({0.25+0.3*cos(45)+0.6*cos(45)*cos(45)},{0.3+0.6*sin(45)*sin(45)});
\draw[densely dashed,thick] ({0.25+0.3*cos(45)+0.6*cos(45)*sin(45)},{0.3+0.6*sin(45)*sin(45)}) arc (-40:240:0.13);

\draw[dashed,thick] (2,0) -- ({0.25+0.3*cos(45)+1.2*cos(45)*cos(45)},{0.3+1.2*sin(45)*sin(45)});
\draw[densely dashed,thick] ({0.25+0.3*cos(45)+1.2*cos(45)*sin(45)},{0.3+1.2*sin(45)*sin(45)}) arc (-40:240:0.13);

\draw (-2,-2) -- (2,2);
\draw (-2,2) -- (2,-2);
\node[rotate=-45] at ({0.15+1.2*cos(135)},{0.15+1.2*sin(135)}) {{\scriptsize{$\leftarrow V=0 \rightarrow$}}};
\node[rotate=45] at ({-0.15+1.2*cos(225)},{0.15+1.2*sin(225)}) {{\scriptsize{$\leftarrow U=0\rightarrow $}}};

\node at (-1.6,-2.3) {{\scriptsize{$V = -\frac{\pi}{2}$}}};
\node at (2.5,2.3) {{\scriptsize{$ V=\frac{\pi}{2} $}}};

\node at (2.5,-2.3) {{\scriptsize{$U = -\frac{\pi}{2}$}}};
\node at (-1.6,2.3) {{\scriptsize{$ U=\frac{\pi}{2} $}}};

\node at (2.6,0) {{\scriptsize{$M_L$}}$\downarrow$};

\node at (0,-3) {$(b)$};

\end{scope}

\begin{scope}[shift={(8,0)},scale=0.8]

\draw[color=blue!100, fill=blue!60, thick] (2,2) -- (2.4,-1.6) -- (0,0);

\draw[color=blue!80, fill=green!40, thick] (2,2) -- (0,0) -- (2,-2);

\draw[dashed,thick] (2,-1.2) -- ({0.25+0.3*cos(45)},0.3);
\draw[densely dashed,thick] ({0.25+0.3*cos(45)},0.3) arc (-40:-330:0.13);

\draw[dashed,thick] (2,-0.6) -- ({0.25+0.3*cos(45)+0.6*cos(45)*cos(45)},{0.3+0.6*sin(45)*sin(45)});
\draw[densely dashed,thick] ({0.25+0.3*cos(45)+0.6*cos(45)*sin(45)},{0.3+0.6*sin(45)*sin(45)}) arc (-40:-330:0.13);

\draw[dashed,thick] (2,0) -- ({0.25+0.3*cos(45)+1.2*cos(45)*cos(45)},{0.3+1.2*sin(45)*sin(45)});
\draw[densely dashed,thick] ({0.25+0.3*cos(45)+1.2*cos(45)*sin(45)},{0.3+1.2*sin(45)*sin(45)}) arc (-40:-330:0.13);

\node[] at (2.5,1) {\scriptsize{\(M_L\)}};
\draw[->] (2.4,0.8)--(2.5,0); %down arrow

\node[] at (1.8,0.7) {\scriptsize{\(M_R\)}};
\draw[->] (1.8,0.9)--(1.8,1.6);

\node at (1.5,-3) {$(c)$};

\end{scope}

\end{tikzpicture}
\end{center}
\caption{
 Holographic SK contour :
In figures  (a) and (b), we show how the spacetimes $M_R$ and $M_L$ can be embedded into Penrose diagrams. Note that the uncoloured parts of the Penrose diagrams are NOT included in the holographic SK contour. In figure (c), we show how these two spacetimes $M_R$ and $M_L$ are glued along their  stretched horizons via a horizon cap. The black dashed lines denote the fixed \(v\) contours shown in fig.\ref{fig:fixedv}.
}
\label{fig.Penrose diagram}
\end{figure}
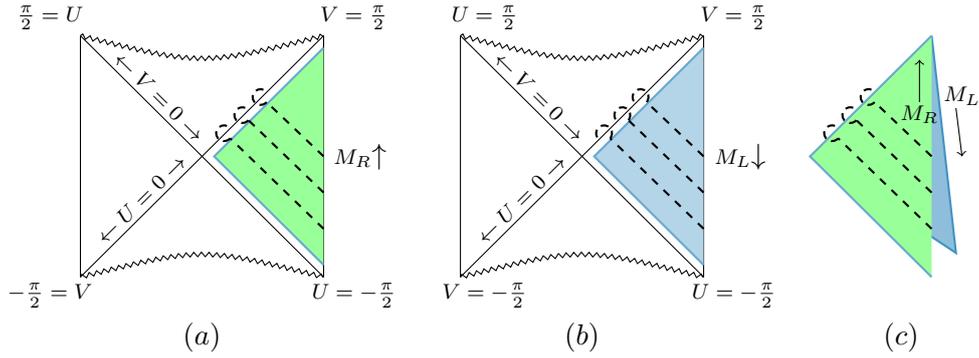

A Penrose diagram of the full holographic SK contour is given in fig.\ref{fig.Penrose diagram}. The coordinates in these diagrams are obtained from $\{v,\rho\}$ by first introducing the outgoing EF coordinate
\begin{equation}
u\equiv v-\frac{2}{r_h} \int_{\infty}^{\rho} \frac{d\rho^\prime}{\rho^{\prime 2} (1-\frac{1}{\rho^{\prime d}})}.
\end{equation}
Notice that the integrand in the above expression of $u$ has a pole at $\rho^\prime=1$. This, on performing the integral, leads to a branch point in $u$ at $\rho=1$. We choose the corresponding branch cut to extend from $\rho=1$ to $\rho\rightarrow \infty$. The values of u on the two sides of the branch cut differ by $i\beta=\frac{4\pi i}{dr_h}$. Therefore, on one hand, the imaginary part of u distinguishes the points on the two patches $M_R$ and $M_L$. On the other hand, the real part of u can replace $\rho$ as a coordinate within each of these patches. Using this real part of  $u$  and the coordinate $v$, we define the following compact coordinates for the Penrose diagrams of $M_R$ and $M_L$ (see fig.\ref{fig.Penrose diagram}):
\begin{equation}
U\equiv -\tan^{-1}\left(e^{-\text{Re}[u] \fr{d r_h}{2}}\right),\ V\equiv \tan^{-1}\left(e^{v \fr{d r_h}{2}}\right).
\end{equation}
The diagram of the full  spacetime can be obtained by gluing these two patches via an analytic continuation of the radial direction near the future horizons.

Let us now be more explicit in describing the geometry along the radial direction. This can be done by extending the definition of the co-ordinate $\xi$ to the entire radial contour. 
One of the primary advantages of working with the $\xi$ co-ordinate is the fact that it can be used to describe the radial contour clearly. We take the convention that $\xi(\rho_c + i \epsilon) = 0$.

Given the differential equation for $\xi$,  an explicit expression for $\xi$ is given by the contour integral 
\bea
\xi  &\equiv & \fr{d}{2 \pi i}\ointctrclockwise^{\rho}_{\rho_c+i\epsilon }
    \fr{y^{d-4}dy}{y^d -1}\ .
\eea
Here the contour integral is performed over counter-clockwise  holographic SK radial contour starting from a cutoff $\rho_c+i\epsilon $  and going towards  $\rho_c-i\epsilon $   shown in the fig.\ref{fig:fixedv}. By looking at the above contour integral representation, we note that the integrand has a simple pole at the horizon with unit residue. It follows that the function $\xi$ has a branch cut with a discontinuity
\bea
\xi(\rho_c + i \epsilon) = 0\ ,\qquad \xi(\rho_c-i\epsilon) = 1 \ .
\eea
Thus, $\xi$ is a good co-ordinate on the radial contour which takes purely imaginary values on $M_L$. The imaginary part logarithmically diverges as we approach the horizon. Its real part then jumps by unity as we traverse the horizon cap region to go from the stretched horizon of $M_L$ to the stretched horizon of $M_R$. Thus, as we move along  $M_R$, it is  $\xi-1$ which is purely imaginary. In what follows, we will  use $\xi$ as a useful parameter to specify where we are on the radial contour.

We will conclude this discussion with the following parenthetical remark :  we can alternately use  the clockwise  holographic SK contour as our defining contour to write
\bea
1-\xi  &\equiv & -\fr{d}{2 \pi i}\ointclockwise^{\rho}_{\rho_c-i\epsilon }
    \fr{y^{d-4}dy}{y^d -1}\ .
\eea
We can then use  $1-\xi$ as an equally good co-ordinate on the doubled radial contour. Apart from the various branches of the holographic SK contour, there are also edges which join those branches together. However, as shown by authors of \cite{Skenderis:2008dh,Skenderis:2008dg}, these edge contributions cancel out in the final answer and hence will be neglected from hereon.

\subsection{The Nambu-Goto action}
We will next describe the non-linear sigma model with target space as the holographic SK spacetime defined in the previous subsection. For simplicity, we shall choose the 
worldsheet co-ordinates to be either one  of $\{v,r\}$ or $\{v,\rho\}$ or $\{v,\xi\}$. We will choose $\{v,\rho\}$ for definiteness, though we will also sometimes find it convenient to 
switch to  $\{v,\xi\}$ instead . In the holographic SK contour, we remind the reader that 
$\rho$ varies from   $\rho_c+i\epsilon $  to  $\rho_c-i\epsilon $. 

The rest of the $(d-1)$ transverse  $x^i$ co-ordinates then become worldsheet fields that  describe the transverse motion of the string. We will denote these worldsheet fields by $\X^i(v,\rho)$. 
The Nambu-Goto action for the string can then  be written as 
\beq
\begin{split}
S_{NG}&=-\frac{1}{2\pi\alpha' }  \int dv\ointctrclockwise d\rho\ \sqrt{-h(X)}\ .
\end{split}
\eeq
Here $h$ represents the determinant of the induced metric in the $\{v,\rho\}$ co-ordinates . The radial integral is over the holographic SK spacetime using the counter-clockwise
radial contour that goes from  $\rho_c+i\epsilon $ (or equivalently $\xi=0$)  to  $\rho_c-i\epsilon $ (or equivalently $\xi=1$)   as shown in the fig.\ref{fig:fixedv}. Using the target space metric
\beq
ds^2= r_h^2\left\{ \frac{2}{r_h}\ dv\ d\rho - \rho^2\left(1-\frac{1}{\rho^d}\right) dv^2+ \rho^2 dx^2_{d-1}\right\}\ ,
\eeq
we can compute the Nambu-Goto action. It is convenient to write down the Nambu-Goto action in a hybrid notation involving both $\rho$ and $\xi$. The conversion between them
is given by the differential equation
\beq
 \frac{d\xi}{d\rho}  \equiv  \fr{d}{2 \pi i}
    \fr{\rho^{d-4}}{\rho^d -1} \ .
\eeq

Using these expressions, the Nambu-Goto action takes the form
\beq
\begin{split}
S_{NG}=&-\frac{r_h}{2\pi\alpha' }  \int dv\ointctrclockwise d\rho \ 
\Bigg[1+\left(\frac{\eta}{r_h}\frac{d\xi}{d\rho}\right)\ \left(\fr{\partial \X^i}{\partial \xi}+i\be \rho^2\fr{\partial  \X^i}{\partial v}\right)\fr{\partial  \X^i}{\partial \xi} \\
&-\frac{1}{4}\left(\frac{\eta}{r_h}\frac{d\xi}{d\rho}\right)^2 \ \sum_{i<j} \left\{\left(\fr{\partial \X^i}{\partial \xi}+i\be \rho^2\fr{\partial  \X^i}{\partial v}\right)\fr{\partial  \X^j}{\partial \xi} -\left(\fr{\partial \X^j}{\partial \xi}+i\be \rho^2\fr{\partial  \X^j}{\partial v}\right)\fr{\partial  \X^i}{\partial \xi} \right\}^2
 \Bigg]^{\fr{1}{2}}\ .
\end{split}
\eeq
Here we have used the notation introduced before in the context of the black brane 
\beq
\beta\equiv\frac{4\pi}{dr_h}  \ , \quad \eta \equiv r_h^3 \frac{d}{2\pi i }\ .
\eeq
This action has to be varied and the resultant differential equations have to be solved with the boundary conditions
\bea
\X^i(v,\rho_c + i \epsilon) = \qL^i(v) \ ,\qquad \X^i(v,\rho_c-i\epsilon) = \qR^i(v) \ .
\eea

In order to derive a tractable set of equations of motion and solve them, we proceed by substituting into this action, an amplitude expansion of the form 
\beq
X^i = X^i_1+X^i_3+\ldots\ ,
\eeq
where we take  $X^i_{2k+1} \sim O(q^{2k+1})$. Note that given the $\X^i\mapsto-\X^i$ symmetry of the equations of motion, under which the boundary conditions flip to $q^i\mapsto-q^i$,
the expansion of $\X^i$ should necessarily be odd in $q^i$. So, no even terms occur in the above expansion. If the boundary displacements of the endpoints are not too large,
we can solve the relevant equations by an amplitude expansion.

As a first step, we get the quadratic action at the second order in amplitude expansion as  
\beq
\begin{split}
S^{(2)}&=-\frac{r_h}{4\pi\alpha' }  \int dv\ointctrclockwise d\rho \left(\frac{\eta}{r_h}\frac{d\xi}{d\rho}\right)\ \left(\fr{\partial \X^i_1}{\partial \xi}+i\be \rho^2\fr{\partial  \X^i_1}{\partial v}\right)\fr{\partial  \X^i_1}{\partial \xi}\ ,
\end{split}
\eeq
where we ignore the contributions of the initial time slice at $v\rightarrow-\infty$.
Varying this action, we get the linearised equation of motion
\beq
\begin{split}
\frac{1}{2} \frac{\partial}{\partial\xi}\left(\fr{\partial }{\partial \xi}+i\be \rho^2\fr{\partial  }{\partial v}\right)\X^i_1+ \frac{1}{2} \left(\fr{\partial }{\partial \xi}+i\be \rho^2\fr{\partial }{\partial v}\right)\frac{\partial \X^i_1}{\partial\xi}
=
\frac{\partial^2\X^i_1}{\partial\xi^2}+i\be  \rho \frac{\partial}{\partial\xi}\left[\rho\fr{\partial  \X^i_1}{\partial v}\right] =0  \ .
\end{split}
\eeq
The appropriate boundary conditions for $\X^i_1$ are given by 
\bea
\X^i_1(v,\rho_c + i \epsilon) = \qL^i(v) \ ,\qquad \X^i_1(v,\rho_c-i\epsilon) = \qR^i(v) \ .
\eea
We note that similarly the boundary conditions for the higher order corrections are given by  
\bea
\X^i_{2k+1}(v,\rho_c + i \epsilon) = 0 \ ,\qquad \X^i_{2k+1}(v,\rho_c-i\epsilon) =0\ .
\eea

We will obtain the solution of this equation  in the subsequent sections. Assuming the solution satisfying appropriate boundary conditions is given, we would be interested in computing 
the on-shell effective action on the solution. This calculation can be simplified using  the following identity established by the use of  linearised equations of motion :
\beq
\begin{split}
\frac{\partial^2}{\partial\xi^2}\left[\frac{1}{2}\X_1^2\right]+ \frac{\partial}{\partial v}\left(i\beta \rho \frac{\partial}{\partial\xi} \left[\frac{\rho}{2}\X_1^2\right]\right)
=\left(\fr{\partial \X_1^i}{\partial \xi}+i\be \rho^2\fr{\partial  \X_1^i}{\partial v}\right)\fr{\partial  \X_1^i}{\partial \xi}  \ .
\end{split}
\eeq
This shows that the quadratic Lagrangian is a total derivative and the whole action can be reduced to boundary terms : 
\beq
\begin{split}
S^{(2)}_{\text{on-shell}}&=-\frac{\eta}{4\pi\alpha' }  \int dv \left[\X_1^i\fr{\partial  \X_1^i}{\partial \xi}\right]_{\xi=0}^{\xi=1}\ .
\end{split}
\label{onshell quadratic}
\eeq

We will now move to the next order in the amplitude expansion. The quartic action has two terms : the first contribution is given by substituting the linearised solution 
to the leading non-linear correction :
\beq
\begin{split}
S^{(4)}&=\frac{r_h}{16\pi\alpha' }  \int dv\ointctrclockwise d\rho \left(\frac{\eta}{r_h}\frac{d\xi}{d\rho}\right)^2\ \left(\fr{\partial \X^i_1}{\partial \xi}+i\be \rho^2\fr{\partial  \X^i_1}{\partial v}\right)\left(\fr{\partial \X^i_1}{\partial \xi}+i\be \rho^2\fr{\partial  \X^i_1}{\partial v}\right)\fr{\partial  \X^j_1}{\partial \xi}\fr{\partial  \X^j_1}{\partial \xi}\ .
\end{split}
\eeq
This is $O(q^4)$ in the amplitude expansion. Another contribution at the same order is given by the substituting the sub-leading solution into the quadratic action :
\beq
\begin{split}
\widehat{S}^{(4)}&=-\frac{r_h}{4\pi\alpha' }  \int dv\ointctrclockwise d\rho \left(\frac{\eta}{r_h}\frac{d\xi}{d\rho}\right)\ \left[\left(\fr{\partial \X^i_1}{\partial \xi}+i\be \rho^2\fr{\partial  \X^i_1}{\partial v}\right)\fr{\partial  \X^i_3}{\partial \xi}+ \left(\fr{\partial \X^i_3}{\partial \xi}+i\be \rho^2\fr{\partial  \X^i_3}{\partial v}\right)\fr{\partial  \X^i_1}{\partial \xi}\right]\ .
\end{split}
\eeq
Using the linearised equation of motion, this term can be  rewritten as a boundary term :
\beq
\begin{split}
\left(\fr{\partial \X^i_1}{\partial \xi}+i\be \rho^2\fr{\partial  \X^i_1}{\partial v}\right)\fr{\partial  \X^i_3}{\partial \xi}+ \left(\fr{\partial \X^i_3}{\partial \xi}+i\be \rho^2\fr{\partial  \X^i_3}{\partial v}\right)\fr{\partial  \X^i_1}{\partial \xi}= &\fr{\partial }{\partial \xi}\left\{ \left(\fr{\partial \X^i_1}{\partial \xi}+i\be \rho^2\fr{\partial  \X^i_1}{\partial v}\right) \X^i_3\right\} \\
&+\left(\fr{\partial }{\partial \xi}+i\be \rho^2\fr{\partial }{\partial v}\right)\left\{  \X^i_3\fr{\partial  \X^i_1}{\partial \xi}\right\}\ .
\end{split}
\eeq
We use this to write $\widehat{S}^{(4)}$ as a purely boundary contribution
\beq
\begin{split}
\widehat{S}^{(4)}&=-\frac{r_h}{4\pi\alpha' }  \int dv\left[  \left(2\fr{\partial \X^i_1}{\partial \xi}+i\be \rho^2\fr{\partial  \X^i_1}{\partial v}\right) \X^i_3 \right]_{\xi=0}^{\xi=1}\ .
\end{split}
\eeq
Since the solution $\X^i_3$ satisfies the boundary conditions
\bea
\X^i_3(v,\rho_c + i \epsilon) = 0 \ ,\qquad \X^i_3(v,\rho_c-i\epsilon) = 0\ ,
\eea
we conclude that $\widehat{S}^{(4)} =0$.

To summarise, to the order we require, it suffices to  obtain the solutions of linearised equations of string motion 
$\X^i_1$ on the holographic SK contour satisfying
\beq
\begin{split}
\frac{\partial^2\X^i_1}{\partial\xi^2}+i\be  \rho \frac{\partial}{\partial\xi}\left[\rho\fr{\partial  \X^i_1}{\partial v}\right] =0  \ , \quad \X^i_1(v,\rho_c + i \epsilon) = \qL^i(v) \ ,\qquad \X^i_1(v,\rho_c-i\epsilon) = \qR^i(v)\ ,
\end{split}
\eeq
and substitute it into the following quadratic and quartic terms of the Nambu-Goto action :
\beq
\begin{split}
S_{\text{on-shell}}
 &=-\frac{r_h}{4\pi\alpha' }  \int dv\ointctrclockwise d\rho \left(\frac{\eta}{r_h}\frac{d\xi}{d\rho}\right)\ \left(\fr{\partial \X^i_1}{\partial \xi}+i\be \rho^2\fr{\partial  \X^i_1}{\partial v}\right)\fr{\partial  \X^i_1}{\partial \xi}\\
 &\quad+\frac{r_h}{16\pi\alpha' }  \int dv\ointctrclockwise d\rho \left(\frac{\eta}{r_h}\frac{d\xi}{d\rho}\right)^2\ \left(\fr{\partial \X^i_1}{\partial \xi}+i\be \rho^2\fr{\partial  \X^i_1}{\partial v}\right)\left(\fr{\partial \X^i_1}{\partial \xi}+i\be \rho^2\fr{\partial  \X^i_1}{\partial v}\right)\fr{\partial  \X^j_1}{\partial \xi}\fr{\partial  \X^j_1}{\partial \xi}\\
&=-\frac{\eta}{4\pi\alpha' }  \int dv \left[\X_1^i\fr{\partial  \X_1^i}{\partial \xi}\right]_{\xi=0}^{\xi=1}\\
&\quad +\frac{r_h}{16\pi\alpha' }  \int dv\ointctrclockwise d\rho \left(\frac{\eta}{r_h}\frac{d\xi}{d\rho}\right)^2\ \left(\fr{\partial \X^i_1}{\partial \xi}+i\be \rho^2\fr{\partial  \X^i_1}{\partial v}\right)\left(\fr{\partial \X^i_1}{\partial \xi}+i\be \rho^2\fr{\partial  \X^i_1}{\partial v}\right)\fr{\partial  \X^j_1}{\partial \xi}\fr{\partial  \X^j_1}{\partial \xi}\ .
\end{split}
\eeq 
Since our focus is on linearised solutions, from hereon we will  drop the subscript on  $\X^i_1$. In what follows, we will find it convenient to write the answers in terms of an effective
`t Hooft coupling 
\beq
\begin{split}
\lambda\equiv \frac{16\pi}{\alpha'}\ .
\end{split}
\eeq 

For example, we can use 
\beq
\begin{split}
 \frac{\eta}{4\pi \alpha'} =- i\frac{\lambda}{2d^2\beta^3} \ , \qquad \frac{\eta^2}{16\pi \alpha' r_h} = -2\pi\frac{\lambda}{2d^3\beta^5}\ ,
\end{split}
\eeq  
to write the action in the following form 
\beq
\begin{split}
S_{\text{on-shell}}
 &=\frac{\lambda}{2d^2\beta^3} \int dv\ointctrclockwise \frac{d\rho}{2\pi} \left(2\pi i\frac{d\xi}{d\rho}\right)\ \left(\fr{\partial \X^i_1}{\partial \xi}+i\be \rho^2\fr{\partial  \X^i_1}{\partial v}\right)\fr{\partial  \X^i_1}{\partial \xi}\\
 &\quad+\frac{\lambda}{2d^3\beta^5}  \int dv\ointctrclockwise \frac{d\rho}{2\pi} \left(2\pi i\frac{d\xi}{d\rho}\right)^2\ \left(\fr{\partial \X^i_1}{\partial \xi}+i\be \rho^2\fr{\partial  \X^i_1}{\partial v}\right)\left(\fr{\partial \X^i_1}{\partial \xi}+i\be \rho^2\fr{\partial  \X^i_1}{\partial v}\right)\fr{\partial  \X^j_1}{\partial \xi}\fr{\partial  \X^j_1}{\partial \xi}\\
&=i\frac{\lambda}{2d^2\beta^3}  \int dv \left[\X_1^i\fr{\partial  \X_1^i}{\partial \xi}\right]_{\xi=0}^{\xi=1}\\ 
&\quad+\frac{\lambda}{2 d^3\beta^5}  \int dv\ointctrclockwise \frac{d\rho}{2\pi} \left(2\pi i\frac{d\xi}{d\rho}\right)^2\ \left(\fr{\partial \X^i_1}{\partial \xi}+i\be \rho^2\fr{\partial  \X^i_1}{\partial v}\right)\left(\fr{\partial \X^i_1}{\partial \xi}+i\be \rho^2\fr{\partial  \X^i_1}{\partial v}\right)\fr{\partial  \X^j_1}{\partial \xi}\fr{\partial  \X^j_1}{\partial \xi}\ .
\end{split}
\eeq

\subsection{Ingoing quasi-normal modes}
Before we try to solve the linearised string equation in the holographic SK contour, we will get our feet wet by solving the familiar problem of ingoing quasi-normal
modes. This does not involve any doubling of spacetime or ideas from Schwinger-Keldysh contour and is a straightforward application of familiar AdS/CFT
techniques. Thus, we will be brief,  relegating the technical details to the appendix \ref{app:DerivExp}. 
 
 We are interested in the solution to the linearised string equation satisfying ingoing boundary conditions at the horizon. Equivalently, our present interest is
in solving for the retarded bulk to boundary Green's function with in-falling boundary conditions at the horizon.

We begin with the linearised string equation of motion 
\beq
\begin{split}
\frac{\partial^2\X^i}{\partial\xi^2}+i\be  \rho \frac{\partial}{\partial\xi}\left[\rho\fr{\partial  \X^i}{\partial v}\right] =0 
\end{split}
\eeq
and seek a solution to it of the form 
\beq
\begin{split}
\X^i=\int dv_0\ q^i(v_0)\ \left[\int\frac{d\w}{2\pi} g_\w(\rho) e^{-i\w(v-v_0)}\right] \ ,
\end{split}
\eeq
where $q^i(v)$ is a given evolution of the string endpoint at the AdS boundary. Here 
$g_\w(\rho)$ is the retarded bulk to boundary Green's function in frequency domain which satisfies 
\beq
\begin{split}
\frac{d^2g_\w}{d\xi^2}+\be \w \rho \frac{d}{d\xi}[\rho g_\w] =0  
\end{split}
\eeq
with the boundary conditions being regularity at the horizon and
\beq
\begin{split}
\lim_{\rho\to\rho_c} g_\w =1\ . 
\end{split}
\eeq
 The regularity at the horizon implies 
 \beq
\begin{split}
\lim_{\rho\to 1}  \frac{dg_\w}{d\xi}=0\ . 
\end{split}
\eeq
Thus, $g_\w$ is the unique solution of the second order ODE quoted above with Dirichlet boundary condition at the AdS boundary ($\rho=\rho_c$) and 
Neumann boundary condition at the horizon ($\rho=1$).

As we show in appendix \ref{app:DerivExp}, we can solve this problem at arbitrary orders in derivative expansion by writing
\beq
g_\w \equiv \exp\left(\sum_{k=1}^\infty (-)^k \left(\frac{\be \w}{2}\right)^k  \psi_k\right)= \exp\left(-\left(\frac{\be \w}{2}\right)\psi_1+\left(\frac{\be \w}{2}\right)^2\psi_2-\left(\frac{\be \w}{2}\right)^3\psi_3+\ldots\right)\ ,
\eeq
with the sequence of radial functions $\psi_k(\rho)$ defined recursively via the integrals 
\beq
\begin{split}
\psi_1  &\equiv \left(\fr{d}{2 \pi i}\right)\int_{\rho_c}^{\rho}     \fr{y^{d-4}dy}{y^d -1}   (y^2-1)\ ,\\
\psi_k &\equiv \left(\fr{d}{2 \pi i} \right)^2 \int_{\rho_c}^{\rho} \fr{y^{d-4} dy}{y^d -1}\int_{1}^{y}  \fr{t^{d-4}dt}{t^d -1}\left[2 \rho^2 \frac{d\psi_{k-1}}{d\xi}-\sum_{m=1}^{k-1}\left(\frac{d\psi_m}{d\xi}\right)\left(\frac{d\psi_{k-m}}{d\xi}\right)\right]_{\rho\to t}\ .
\end{split}
\eeq
Note that $\frac{d\psi_1}{d\xi}=\rho^2-1$ with the consequence that $\lim_{\rho\to 1}  \frac{d\psi_1}{d\xi}=0$. 

Further, if we assume that $\lim_{\rho\to 1}  \frac{d\psi_k}{d\xi}=0$ for all $k<k_0$, then  the integrand for $\psi_{k_0}$ has no horizon poles and is hence analytic in the ingoing Eddington Finkelstein co-ordinates. This implies that  $\lim_{\rho\to 1}\frac{d\psi_{k_0}}{d\xi}=0$. Thus, by induction, all $\psi_k(\rho)$ and in turn, $g_\w$ are analytic in the ingoing Eddington Finkelstein co-ordinates. The properties
\beq
\begin{split}
\lim_{\rho\to\rho_c} \psi_k =0\ , \quad
\lim_{\rho\to 1}  \frac{d\psi_k}{d\xi}=0\ ,
\end{split}
\eeq
guarantee that $g_\w$ satisfies the correct boundary conditions.

In the special case of $d=2$, i.e., in AdS$_3$, the sequence of radial functions $\psi_k(\rho)$  take an especially simple  form
\bea
\psi_k[\text{AdS}_3]=\fr{1}{k}\left(\fr{d}{2\pi i}\right)^k \left( \fr{1}{\rho_c^k}-\fr{1}{\rho^k}\right)\ .  
\eea
These can be summed to get a simple closed form expression for the retarded Green's function
\bea
g_\w[\text{AdS}_3]=\left(\fr{\rho_c}{\rho}\right)\fr{\w+i\ r_h \rho}{\w+i\ r_h \rho_c}\ .
\eea
These AdS$_3$ expressions nicely illustrate the general properties argued above.

Once the causal bulk to boundary Green's function with in-falling boundary conditions has been constructed, the ingoing solution for
the string with a given boundary evolution $q^i(v)$ is given by the convolution integral
\beq
\begin{split}
\X^i_{\text{ingoing}}=\int dv_0\ q^i(v_0)\ \left[\int\frac{d\w}{2\pi} g_\w e^{-i\w(v-v_0)}\right] =\int\frac{d\w}{2\pi} g_\w q^i(\w) e^{-i\w v} \ .
\end{split}
\eeq
In frequency domain, we simply have $\X^i(\w)= g_\w q^i(\w)$.
The expression above describes the linear retarded response of the string to the motion of its endpoint. It also gives a linearised  description of how the excitations on the string 
fall into the black brane and get dissipated away. 

\subsection{Outgoing Hawking modes}
We will now  shift to a discussion of the \emph{other} solution to the equation which is \emph{not} ingoing. These are needed to describe Hawking radiation
of the string modes and their effects. As we shall see, a correct IR divergence-free description of these modes should really be done within the holographic
Schwinger-Keldysh contour with doubled spacetime rather than in the single copy spacetime. Nevertheless, it is useful to set the stage by describing these
modes in the language of the  single copy spacetime, which we will do in this subsection.

A naive attempt at getting the outgoing modes from the ingoing modes would be to just reverse the time argument. Given that the ingoing solution of frequency  $\w$ has the 
form $g_\w e^{-i\w v}$, the naive time reversed solution would be  $g_\w e^{i\w v}$. But this does not quite work, since the string 
equations of motion in the ingoing Eddington-Finkelstein co-ordinates is
\beq
\begin{split}
\frac{\partial^2\X^i}{\partial\xi^2}+i\be  \rho \frac{\partial}{\partial\xi}\left[\rho\fr{\partial  \X^i}{\partial v}\right] =0 
\end{split}
\eeq
and it is \emph{not} invariant under $v\mapsto -v$. 

The customary solution to this impasse is to pass to a co-ordinate system (say Kruskal co-ordinates) where the underlying time-reversal invariance becomes manifest. Here, we 
will resist the temptation to do so for two reasons: one, it is not clear such a solution generalises  to the fully dynamical case where black brane itself has 
formed out of a collapsing shell of matter in the past. The second reason is as follows : we take the viewpoint that the hiding/spontaneous breaking of time-reversal invariance is in fact
a physical fact of Brownian motion as it emerges from the microscopic dynamics. We prefer to keep this physics manifest rather than camouflaging it.

How then do we deal with the hidden/spontaneously broken time reversal invariance in this system ? We will characterise it by an \emph{order parameter} field for time-reversal 
symmetry breaking, which we will denote by the letter $\chi$. We will demand that time reversal invariance be non-linearly realised in our system via the map
\beq
\begin{split}
v\mapsto i\beta \chi-v\ .
\end{split}
\eeq
The normalisation factor of $i\beta$ here is chosen for notational convenience. 

In the context of the solutions to linearised string equations, this practically amounts to demanding that $g_\w e^{\be\w\chi} e^{i\w v}$ be the time-reversed counterpart of the  ingoing solution.
Substituting the ansatz $\X\sim g_\w e^{\be\w\chi} e^{i\w v}$ into the equation above and using the fact that  $g_\w$ satisfies the differential equation 
\beq
\frac{d^2g_\w}{d\xi^2}+\be \w \rho \frac{d}{d\xi}[\rho g_\w] =0\ ,  
\eeq
we get a  differential equation for $\chi$ of the form
\beq
\fr{d}{d\xi}\left[g^2_\w e^{\be\w\chi}\left(\fr{d\chi}{d\xi}-\rho^2\right)\right]=0\ .  
\eeq
We already know the ingoing mode analytic at the future horizon to be $\X\sim g_\w e^{\be\w\chi} e^{i\w v}=g_{-\w}e^{i\w v}$, which gives one solution for $\chi$ in the above equation.
The other solution can be simply obtained by setting 
\beq
\fr{d\chi}{d\xi}=\rho^2=1+\frac{d\psi_1}{d\xi} \ .  
\eeq
Here we have used the fact that $\frac{d\psi_1}{d\xi} =\rho^2-1$, where $\psi_1$ is the function that appeared in the last subsection as the one that governs the retarded response of the string to leading order in small $\w$ expansion. Thus, we conclude that if we take $\chi=\xi+\psi_1$ , then $g_\w e^{\be\w\chi} e^{i\w v}$ is the required time-reversed solution.

Given that the retarded Green's function takes the form
\beq
\begin{split}
g_\w e^{-i\w v} \equiv & \exp\left(\sum_{k=1}^\infty (-)^k \left(\frac{\be \w}{2}\right)^k  \psi_k\right)e^{-i\w v}\\
=& \exp\left(-\left(\frac{\be \w}{2}\right)\psi_1+\left(\frac{\be \w}{2}\right)^2\psi_2-\left(\frac{\be \w}{2}\right)^3\psi_3+\ldots\right)e^{-i\w v}\ ,
\end{split}
\eeq
its time-reversed counterpart is given by 
\beq
g_\w e^{\be\w\chi} e^{i\w v} = \exp\left(-\left(\frac{\be \w}{2}\right)(-2\xi-\psi_1)+\left(\frac{\be \w}{2}\right)^2\psi_2-\left(\frac{\be \w}{2}\right)^3\psi_3+\ldots\right)e^{i\w v}\ .
\eeq
Thus, time reversal at the level of linearised solutions is achieved by a replacement $\psi_1\mapsto (-2\xi-\psi_1)$ keeping all other $\psi_k$'s intact.

Another perspective on $\chi$ is obtained by writing its differential equation in terms of $\rho$ derivative:
\beq
\fr{d\chi}{d\rho}=\frac{d}{2\pi i} \frac{1}{\rho^2\left(1-\frac{1}{\rho^d}\right)}\ .  
\eeq
We recognise here the inverse of the black brane redshift factor which then identifies $\chi$ as the tortoise co-ordinate up to a pre-factor. A cognizant reader might have
realised that this is to be expected : from the point of view of Kruskal co-ordinates, the time reversed solution that we have constructed is just $g_\w e^{i\w u}$ with
$u$ being the outgoing Eddington Finkelstein co-ordinate. Converting $u$ to $v$ gives a shift proportional to tortoise co-ordinate which explains the origin of $\chi$
from the Kruskal viewpoint.
\subsection{The doubled string on holographic SK contour}
We will now put together the above two solutions to determine the configuration that satisfies the correct boundary conditions. 
We begin with the most general solution of the linearised string equations, i.e., 
\beq
\begin{split}
\X^i=\int\frac{d\w}{2\pi}\ [\ c^i_\w g_\w+h^i_\w g_{-\w} e^{-\be\w\chi}]\  e^{-i\w v} \ ,
\end{split}
\eeq
where we have redefined $\omega$ in the Hawking mode to have the standard sign for Fourier transform. In frequency domain, we 
can simply write 
\beq
\begin{split}
\X^i(\w,\xi)= c^i_\w g_\w(\xi)+h^i_\w g_{-\w}(\xi)\ e^{-\be\w\chi(\xi)} \ .
\end{split}
\eeq
This solution has a causal Green's function part $c_\w$ which represents the  in-falling quasi-normal mode, superposed with the Hawking mode $h_\w$.
While the ingoing quasi-normal modes are regular and give a smooth description of the string dynamics, this is not true about outgoing Hawking modes
which are divergent at the horizon due to the $\chi$ function. Their physical import is unclear a priori, nor is it evident how to extract finite answers from them staying within outgoing
EF co-ordinates. As we will now argue, the correct interpretation of the Hawking modes is that they appear as saddle points in the Schwinger-Keldysh
path integral of the string world-sheet.

We can embed the string solution into our holographic SK contour by interpreting the $\xi$'s in the above expression as varying over the holographic SK
contour, i.e., the $\xi$ function begins from zero in the left contour boundary (at $\rho=\rho_c+i\varepsilon$) and ends in becoming unity in the right 
contour boundary (at $\rho=\rho_c-i\varepsilon$) after traversing the contour. This is a simple lifting of the one copy solution to a configuration in the
entire holographic SK contour, whereby the problematic near horizon region gets excised. The $\xi$ and hence the $\chi$ function are now
completely regular on the new contour. 

One can now interpret the complex spacetime, i.e., the horizon cap region between two stretched horizons 
as the geometry which captures the Hawking radiation process within Schwinger-Keldysh formalism. Since
Hawking radiation  can often be thought of as a tunnelling  process (see \cite{Vanzo:2011wq}
and references therein), one can interpret the string configuration within the horizon cap region as a `Schwinger-Keldysh instanton'
that computes the corresponding tunnelling probability. This perspective is familiar in the activation type problems dealt within
Schwinger-Keldysh formalism \cite{Caldeira:1982uj,kamenev_2011}. Such saddle points with the difference fields turned on, are
crucial in capturing the physics of dissipative tunnelling processes, It is thus interesting that this tunnelling perspective of Hawking
modes emerges naturally out of holographic SK formallism.

Let us return back to our solution to impose the appropriate boundary conditions. We would like to impose
\bea
\X^i(\rho_c + i \epsilon) = \qL^i \ ,\qquad \X^i(\rho_c-i\epsilon) = \qR^i \ .
\eea
We need $g_{\w}(\rho_c \pm i \epsilon)=1$ along with
\beq
\begin{split}
\chi(\rho_c+i \epsilon) = \xi(\rho_c+i \epsilon)+\psi_1(\rho_c+i \epsilon)=\xi(\rho_c+i \epsilon)=0\ 
\end{split}
\eeq
and
\beq
\begin{split}
\chi(\rho_c-i \epsilon) = \xi(\rho_c-i \epsilon)+\psi_1(\rho_c-i \epsilon)=\xi(\rho_c-i \epsilon)=1\ .
\end{split}
\eeq
This yields the equations
\beq
\begin{split}
c^i_\w +  h^i_\w &= \qL^i(\w)\ , \\  
c^i_\w + e^{-\be\w} h^i_\w &= \qR^i(\w) \ 
\end{split}
\eeq
in the frequency domain. These are readily solved to give the  Son-Teaney doubled string configuration \cite{Son:2009vu} 
\beq
\X^i(\w) = g_\w [(1+f_\w) \qR^i(\w) -f_\w \qL^i(\w) ] - g_{-\w} e^{-\be \w \chi} (1+f_\w)(\qR^i(\w) -\qL^i(\w) )\ .
\eeq
Here, $f_\w$ is  the  Bose-Einstein factor given by 
\beq
f_\w \equiv \frac{1}{e^{\be\w}-1}\ .
\eeq
The Bose-Einstein factor $f_\w$ satisfies $1+f_\w=e^{\be \w}f_\w$ and $1+f_\w+f_{-\w}=0$. Using this, the solution above can also be written equivalently as
\beq
\X^i(\w) = g_\w [(1+f_\w) \qR^i(\w) -f_\w \qL^i(\w) ] - g_{-\w} e^{\be \w (1-\chi)} f_\w(\qR^i(\w) -\qL^i(\w) )\ .
\eeq

The solution satisfies a reality condition imposed by demanding invariance under the exchanges 
\beq
\qR^i \leftrightarrow \qL^i \ , \quad \w \rightarrow -\w\ , \quad i \rightarrow - i\ .
\eeq 
Thus, we demand that $[\X^i(-\w)]^\ast$ takes the same form as $\X^i(\w)$ with an exchange of the right/left boundary value data.
We have (using $1+f_\w=e^{\be \w}f_\w$ and $1+f_\w+f_{-\w}=0$)
\beq
\X^i(-\w) = g_{-\w} [(1+f_\w) \qL^i(-\w)-f_\w \qR^i(-\w) ] -g_{\w} e^{-\be \w (1-\chi)} (1+f_\w)(\qL^i(-\w) -\qR^i(-\w) )\ .
\eeq
Since, $\qR^i,\qL^i$ are real functions in time domain, they satisfy  $[q(-\w)]^\ast =q(\w)$. For the same reason, we have $[g_{-\w}]^\ast =g_\w$
for the retarded Green's function.  Complex conjugating the holographic SK contour maps $\xi$ to $1-\xi^\ast$ and hence $\chi$ to $1-\chi^\ast$.
With these, it follows that $[\X^i(-\w)]^\ast$ takes the same form as $\X^i(\w)$ after a complex conjugation and right/left exchange.

\section{Non-linear Langevin theory}\label{sec:nonlinLang}
 Before proceeding to use this configuration to compute the influence phase, it is useful to define the following combinations
that recur in our solutions :
\beq
\begin{split}
\qbF^i (\w)&\equiv f_{-\w} \qR^i(\w) -(1+f_{-\w}) \qL^i(\w)=-(1+ f_{\w}) \qR^i(\w) +f_{\w} \qL^i(\w)\ , \\
\qbP^i (\w) &\equiv (1+f_{-\w}) \left(\qR^i(\w) - \qL^i(\w)\right)=-f_{\w} \left(\qR^i(\w) - \qL^i(\w)\right)\ .
\end{split}
\eeq
These combinations correspond to the retarded advanced (RA) basis within the SK formalism \cite{Hou:1998yc,Wang:1998wg,Henning:1993gh,Chu:1993nc,Henning:1995sm,Chaudhuri:2018ymp,Baier:1993yh,Aurenche:1991hi,vanEijck:1992mq,vanEijck:1994rw} which is the most convenient
in making KMS conditions manifest. In terms of these combinations, we can then write
\beq
\begin{split}
\X^i= -g_\w \qbF^i(\w)+g_{-\w} e^{\be\w(1-\chi)} \qbP^i (\w)\ .
\end{split}
\eeq
Thus, we see that RA basis combinations appear naturally in holography :  $\qbF^i$ sources the ingoing quasi-normal mode regular at the future horizon whereas $\qbP^i$ excites the Hawking mode.

\subsection{Influence phase of the heavy quark}\label{sec:IP of heavy quark}
We will now compute the influence phase of the heavy quark in the retarded-advanced basis. To this end, it is  useful to define the derivative combinations
\beq
\begin{split}
(-i)\left[\fr{d \X^i}{d\xi}+\be\w \rho^2\X^i \right] &= -A_\w \qbF^i(\w)+ B_{-\w} e^{\be\w(1-\chi)} \qbP^i (\w)\ ,\\ (-i)\fr{d \X^i}{d\xi} &=-B_{\w} \qbF^i (\w)+A_{-\w} e^{\be\w(1-\chi)} \qbP^i (\w)\ .
\end{split}
\eeq
where 
\beq
\begin{split}
iA_\w &\equiv \fr{d g_\w}{d\xi}+\be\w \rho^2g_\w=e^{\be\w(1-\chi)}\fr{d}{d\xi} [g_\w e^{-\be\w(1-\chi)}]\ , \\
iB_\w &\equiv \fr{d g_\w}{d\xi} =e^{\be\w(1-\chi)}\left\{\fr{d}{d\xi} [g_\w e^{-\be\w(1-\chi)}]-\be\w \rho^2g_\w e^{-\be\w(1-\chi)}\right\}\ .
\end{split}
\eeq
Here $A_\w, B_\w$ are analytic retarded functions which are well-behaved  at the horizon with $B_\w$ vanishing at the horizon.

With these definitions, we can compute the influence phase of the heavy quark as 
\beq
\begin{split}
S^{(2)}_{I}&= i \int \frac{d^2\w}{(2\pi)^2}\  \mathfrak{G}[PF] \qbP^i (\w_1) \qbF^i (\w_2)\  \ ,\\
S^{(4)}_{I}&= i^3 \int \frac{d^4\w}{(2\pi)^4} \left\{  \mathfrak{G}[PFFF]\  \qbP^i (\w_1) \qbF^i (\w_2)\  \qbF^j(\w_3)\ \qbF^j (\w_4)\right. \\ 
  &\qquad \qquad + \mathfrak{G}[PPPF]\  \qbP^i (\w_1) \qbP^i (\w_2)\  \qbP^j(\w_3)\ \qbF^j (\w_4) \\
  &\qquad \qquad  +  \mathfrak{G}[PFPF]\  \qbP^i (\w_1) \qbF^i (\w_2)\  \qbP^j(\w_3)\ \qbF^j (\w_4) \\
  &\qquad \qquad +  \left. \mathfrak{G}[PPFF]\  \qbP^i (\w_1) \qbP^i (\w_2)\  \qbF^j(\w_3)\ \qbF^j (\w_4)\right\}\ ,
\end{split}
\eeq
where \(d^n \w \) stands for \(\prod_{i=1}^{n}d\w_i \) . The functions that appear in the above influence phase evaluate to 
\beq
\begin{split}
 \mathfrak{G}[PF] &\equiv \frac{\lambda}{2d^2\beta^3} 2\pi \ \delta(\sum\w) \ \ointctrclockwise \frac{d\rho}{2\pi i} \left(2\pi i\frac{d\xi}{d\rho}\right)\ 
[A_{-\w_1} A_{\w_2}+B_{-\w_1} B_{\w_2}]\ e^{\be\w_1(1-\chi)}\ ,\\
 \mathfrak{G}[PFFF] &\equiv 2\frac{\lambda}{2d^3\beta^5}\ 2\pi \ \delta(\sum\w) \ \ointctrclockwise  \frac{d\rho}{2\pi i} \left(2\pi i\frac{d\xi}{d\rho}\right)^2  [ B_{-\w_1} A_{\w_2}B_{\w_3}B_{\w_4}\\
 &\quad + A_{-\w_1}B_{\w_2}A_{\w_3}A_{\w_4}] e^{\be\w_1(1-\chi)} \ ,\\
\mathfrak{G}[PPPF] &\equiv 2\frac{\lambda}{2d^3\beta^5}\ 2\pi \ \delta(\sum\w) \ \ointctrclockwise  \frac{d\rho}{2\pi i} \left(2\pi i\frac{d\xi}{d\rho}\right)^2  [ A_{-\w_1} A_{-\w_2}B_{-\w_3} A_{\w_4} \\
&\quad+ B_{-\w_1} B_{-\w_2}A_{-\w_3} B_{\w_4} ]\ e^{-\beta\w_4 (1-\chi)} \ ,\\
\mathfrak{G}[PFPF] &\equiv-2\frac{\lambda}{2d^3\beta^5}\ 2\pi \ \delta(\sum\w) \ \ointctrclockwise  \frac{d\rho}{2\pi i} \left(2\pi i\frac{d\xi}{d\rho}\right)^2\  [A_{-\w_1} B_{\w_2}B_{-\w_3} A_{\w_4} \\
&\quad +B_{-\w_1} A_{\w_2}A_{-\w_3}B_{\w_4}]\ e^{\be(\w_1+\w_3)(1-\chi)} \ ,\\
\mathfrak{G}[PPFF] &\equiv -\frac{\lambda}{2d^3\beta^5}\ 2\pi\ \delta(\sum\w) \ointctrclockwise \frac{d\rho }{2\pi i}  \left(2\pi i\frac{d\xi}{d\rho}\right)^2   [A_{-\w_1} A_{-\w_2} A_{\w_3} A_{\w_4}\\
&\quad +B_{-\w_1} B_{-\w_2} B_{\w_3} B_{\w_4}]\ e^{\be(\w_1+\w_2)(1-\chi)} \ .
\end{split}
\eeq

These expressions describe the variety of  four mode interactions on the world sheet which contribute to the influence phase. Here, the modes with $-\w$ in their subscripts are
the Hawking modes and those with $+\w$ subscript are ingoing.

We note the remarkable fact that the terms of the form $PP$ or $FF$ do not appear in the quadratic influence phase, nor do  the terms of the form $PPPP$ or $FFFF$ appear in the quartic influence phase. To understand this, we begin by looking at the PP and FF part of the quadratic influence phase  which takes the form
\beq
\begin{split}
-i\frac{\lambda}{2d^2\beta^3}  \int \frac{d^2\w}{(2\pi)^2}2\pi\delta(\sum\w)\ointctrclockwise &\frac{d\rho}{2\pi i} \left(2\pi i\frac{d\xi}{d\rho}\right)\\
&\ \left\{
A_{\w_1} B_{\w_2}\ \qbF^i (\w_1) \qbF^i (\w_2) + A_{-\w_1}B_{-\w_2} \ \qbP^i (\w_1) \qbP^i (\w_2)
\right\}\ ,
\end{split}
\eeq
where we have used the identity $e^{(1-\chi)\be\sum \w}=1$. Because of the cancellation of $\chi$, the entire integrand here has the same value in both left and right contours.
Further, the simple pole of 
\[ \left(\frac{d\xi}{d\rho}\right) =\frac{d}{2\pi i}\frac{\rho^{d-4}}{\rho^d-1} \] 
at the horizon $\rho^2-1$ is cancelled by the zero of the  $B_{\w}$ function which vanishes at the horizon. As a result, the integrand is in fact entirely analytic 
and the contributions from two contours cancel each other. A similar cancellation of contours lies behind the vanishing of  $PPPP$ or $FFFF$ terms.

The  $FFFF$ and $PPPP$ part of the quartic influence phase is given by 
\beq
\begin{split}
\frac{i\lambda}{2d^3\beta^5} &\int \frac{d^4\w}{(2\pi)^4}2\pi\delta(\sum\w)\ointctrclockwise  \frac{d\rho}{2\pi i} \left(2\pi i\frac{d\xi}{d\rho}\right)^2 \bigg\{ A_{\w_1} A_{\w_2} B_{\w_3} B_{\w_4}\ \qbF^i (\w_1)\ \qbF^i (\w_2)\  \qbF^j(\w_3)\ \qbF^j (\w_4) \\
 & +A_{-\w_1} A_{-\w_2} B_{-\w_3} B_{-\w_4} \  \qbP^i (\w_1)\ \qbP^i (\w_2)\  \qbP^j(\w_3)\ \qbP^j (\w_4)\bigg\} \ .
\end{split}
\eeq

In this case, we again have a single valued function whose double pole $\left(\frac{d\xi}{d\rho}\right)^2$ at the horizon is cancelled by the zeroes of the two $B_{\w}$ functions at the horizon. As a result, the integrand is entirely analytic and this contribution again cancels between the two contours.

The vanishing of the all $P$ and all $F$ terms is the hallmark of the retarded advanced basis in the Schwinger-Keldysh formalism \cite{Chaudhuri:2018ymp,Baier:1993yh,Aurenche:1991hi,vanEijck:1992mq,vanEijck:1994rw}. On the CFT side, 
this is a combined consequence of both  the largest time equation of SK formalism and the KMS conditions arising from the thermality of the bath. 
It is satisfying to see that this fact appears in our analysis based on general grounds of analyticity, independent of the underlying details. 

On the gravity side, we see that all the contributions to the influence phase arise from interactions between the ingoing modes and the Hawking modes. Both the self-interaction between the ingoing modes as well as the self-interaction effects of Hawking modes cancel out in the influence phase. This, perhaps, is an indication that on the CFT side the dominant physics we are looking
at is the interaction between the quark and the CFT bath rather than effects arising from self-interactions. Further the influence phase, in general, receives contributions from the whole 
of the exterior spacetime (due to the branch cut in $\chi$ function along the radial direction). However, as we shall see below, at leading order in low frequency/derivative expansion, we get localised contributions from the horizon.

\subsection{Influence phase in Keldysh basis and its derivative expansion}
While the influence phase stated in terms of the retarded advanced basis $\qbP^i$ and $\qbF^i$ is very efficient at exhibiting the essential thermal structure of the influence phase,
it is not convenient for studying the low frequency response. To this end,  we will begin here  by making the Bose-Einstein factors manifest so as to study the derivative expansion.
This is achieved by substituting 
\beq
\begin{split}
\qbF^i (\w)&=-\qa^i(\w) -N_\w \qd^i(\w) \ ,\quad
\qbP^i (\w) =-f_{\w} \qd^i(\w) \ ,
\end{split}
\eeq
where 
\[ N_\w\equiv f_{\w}+\frac{1}{2}\]
 in the expressions for the influence phase. Here $\qa^i\equiv \fr{1}{2}(\qR^i+\qL^i)$ and $\qd^i\equiv \qR^i-\qL^i$ constitute the Keldysh
or average-difference basis \cite{Keldysh:1964ud,CHOU19851,kamenev_2011}. As we did in the retarded-advanced basis, we will find it convenient to define \
\beq
\begin{split}
(-i)\Big(\frac{dX^i}{d\xi}+\beta \omega\rho^2X^i\Big)&=A_\omega \qa^i(\omega)+\widetilde{A}_\omega\qd^i(\omega),\\
 (-i)\frac{dX^i}{d\xi}&=B_\omega \qa^i(\omega)+\widetilde{B}_\omega\qd^i(\omega)\ ,
\end{split}
\eeq
where 
\beq
\begin{split}
\widetilde{A}_\omega\equiv N_\omega A_\omega -e^{\beta\omega(1-\chi)}f_\omega B_{-\omega}\ ,\\
\widetilde{B}_\omega\equiv N_\omega B_\omega -e^{\beta\omega(1-\chi)}f_\omega A_{-\omega}\ .
\end{split}
\eeq

The influence phase in this basis takes the form
\beq
\begin{split}
S^{(2)}_{I}&= i \int \frac{d^2\w}{(2\pi)^2}\ \left(\mathfrak{G}[d^2]\qd^i(\w_1) \qd^i(\w_2) +  \mathfrak{G}[da]\qd^i(\w_1) \qa^i(\w_2)+  \mathfrak{G}[a^2]\qa^i(\w_1) \qa^i(\w_2)\right)  \ ,\\
S^{(4)}_{I}&= i^3 \int \frac{d^4\w}{(2\pi)^4} \Big\{  \mathfrak{G}[d^4]\  \qd^i (\w_1) \qd^i (\w_2)\  \qd^j(\w_3)\ \qd^j (\w_4) 
+ \mathfrak{G}[d^3a]\  \qd^i (\w_1) \qd^i (\w_2)\  \qd^j(\w_3)\ \qa^j (\w_4)  \\
&\qquad  +   \mathfrak{G}[dada]\  \qd^i (\w_1) \qa^i (\w_2)\  \qd^j(\w_3)\ \qa^j (\w_4)+   \mathfrak{G}[d^2a^2]\  \qd^i (\w_1) \qd^i (\w_2)\  \qa^j(\w_3)\ \qa^j (\w_4) \\
&\qquad  +   \mathfrak{G}[da^3]\  \qd^i (\w_1) \qa^i (\w_2)\  \qa^j(\w_3)\ \qa^j (\w_4)+   \mathfrak{G}[a^4]\  \qa^i (\w_1) \qa^i (\w_2)\  \qa^j(\w_3)\ \qa^j (\w_4)\Big\} \ ,
\end{split}
\eeq
where we have defined the coefficient functions
\beq
\begin{split}
&\mathfrak{G}[d^2]=-\frac{1}{2}\frac{\lambda}{2d^2\beta^3}2\pi\delta(\sum \omega)\ointctrclockwise\frac{d\rho}{2\pi i}\Big(2\pi i\frac{d\xi}{d\rho}\Big)\Big(\widetilde{A}_{\omega_1}\widetilde{B}_{\omega_2}+\widetilde{B}_{\omega_1}\widetilde{A}_{\omega_2}\Big)\ ,\\
&\mathfrak{G}[da]=-\frac{\lambda}{2d^2\beta^3}2\pi\delta(\sum \omega)\ointctrclockwise\frac{d\rho}{2\pi i}\Big(2\pi i\frac{d\xi}{d\rho}\Big)\Big(\widetilde{A}_{\omega_1}B_{\omega_2}+\widetilde{B}_{\omega_1}A_{\omega_2}\Big)\ ,\\
&\mathfrak{G}[a^2]=-\frac{1}{2}\frac{\lambda}{2d^2\beta^3}2\pi\delta(\sum \omega)\ointctrclockwise\frac{d\rho}{2\pi i}\Big(2\pi i\frac{d\xi}{d\rho}\Big)\Big(A_{\omega_1}B_{\omega_2}+B_{\omega_1}A_{\omega_2}\Big)\ ,\\
&\mathfrak{G}[d^4]=-\frac{1}{2}\frac{\lambda}{2d^3\beta^5}2\pi\delta(\sum \omega)\ointctrclockwise\frac{d\rho}{2\pi i}\Big(2\pi i\frac{d\xi}{d\rho}\Big)^2\Big(\widetilde{A}_{\omega_1}\widetilde{A}_{\omega_2}\widetilde{B}_{\omega_3}\widetilde{B}_{\omega_4}+\widetilde{B}_{\omega_1}\widetilde{B}_{\omega_2}\widetilde{A}_{\omega_3}\widetilde{A}_{\omega_4}\Big)\ ,\\
&\mathfrak{G}[d^3a]=-2\frac{\lambda}{2d^3\beta^5}2\pi\delta(\sum \omega)\ointctrclockwise\frac{d\rho}{2\pi i}\Big(2\pi i\frac{d\xi}{d\rho}\Big)^2\Big(\widetilde{A}_{\omega_1}\widetilde{A}_{\omega_2}\widetilde{B}_{\omega_3}B_{\omega_4}+\widetilde{B}_{\omega_1}\widetilde{B}_{\omega_2}\widetilde{A}_{\omega_3}A_{\omega_4}\Big)\ ,\\
&\mathfrak{G}[d^2a^2]=-\frac{\lambda}{2d^3\beta^5}2\pi\delta(\sum \omega)\ointctrclockwise\frac{d\rho}{2\pi i}\Big(2\pi i\frac{d\xi}{d\rho}\Big)^2\Big(\widetilde{A}_{\omega_1}\widetilde{A}_{\omega_2}B_{\omega_3}B_{\omega_4}+\widetilde{B}_{\omega_1}\widetilde{B}_{\omega_2}A_{\omega_3}A_{\omega_4}\Big)\ ,\\
&\mathfrak{G}[dada]=-2\frac{\lambda}{2d^3\beta^5}2\pi\delta(\sum \omega)\ointctrclockwise\frac{d\rho}{2\pi i}\Big(2\pi i\frac{d\xi}{d\rho}\Big)^2\Big(\widetilde{A}_{\omega_1}A_{\omega_2}\widetilde{B}_{\omega_3}B_{\omega_4}+\widetilde{B}_{\omega_1}B_{\omega_2}\widetilde{A}_{\omega_3}A_{\omega_4}\Big)\ ,\\
&\mathfrak{G}[da^3]=-2\frac{\lambda}{2d^3\beta^5}2\pi\delta(\sum \omega)\ointctrclockwise\frac{d\rho}{2\pi i}\Big(2\pi i\frac{d\xi}{d\rho}\Big)^2\Big(\widetilde{A}_{\omega_1}A_{\omega_2}B_{\omega_3}B_{\omega_4}+\widetilde{B}_{\omega_1}B_{\omega_2}A_{\omega_3}A_{\omega_4}\Big)\ ,\\
&\mathfrak{G}[a^4]=-\frac{1}{2}\frac{\lambda}{2d^3\beta^5}2\pi\delta(\sum \omega)\ointctrclockwise\frac{d\rho}{2\pi i}\Big(2\pi i\frac{d\xi}{d\rho}\Big)^2 \Big(A_{\omega_1}A_{\omega_2}B_{\omega_3}B_{\omega_4}+B_{\omega_1}B_{\omega_2}A_{\omega_3}A_{\omega_4}\Big)\ ,\\
\end{split}
\eeq

The coefficient functions $\mathfrak{G}[a^2]$ and $\mathfrak{G}[a^4]$ vanish by arguments  analogous to those given for the vanishing of $\mathfrak{G}[PP]$, $\mathfrak{G}[FF]$, $\mathfrak{G}[PPPP]$ and $\mathfrak{G}[FFFF]$ in the previous subsection. The absence of the corresponding terms in the influence phase implies that the master equation that evolves the quark's density matrix has a Lindblad form \cite{Avinash:2017asn,kamenev_2011}.

We will now like to extract the low frequency behaviour from the full influence phase.
Using the fact that $f_\w,N_\w\sim \fr{1}{\be\w}$, one can see that the low frequency expansions of the functions $ A_\w,\  B_\w,\ \widetilde{A}_\w$ and $\widetilde{B}_\w$ take the following forms
 \beq
\begin{split}
 A_\w &\approx -i \fr{\be\w}{2}(\rho^2 +1) + O(\w^2)\ ,\quad B_\w \approx i\fr{\be\w}{2}(\rho^2 -1)+ O(\w^2)\ ,\\
\widetilde{A}_\w &\approx -i  + O(\w)\ ,\quad  \widetilde{B}_\w \approx -i+ O(\w)\ .
\end{split}
\eeq
 Substituting the above expressions in the quartic coefficient functions, we conclude that their low frequency behaviour is given by
\beq
\begin{split}
 \mathfrak{G}[d^4] & \sim O(\w^0)\ ,\quad \mathfrak{G}[d^3a] \sim O(\w)\ ,\quad \mathfrak{G}[dada]\sim O(\w^2)\ ,\\
 \mathfrak{G}[d^2a^2] & \sim O(\w^2)\ ,\quad  \mathfrak{G}[da^3]\sim  O(\w^3)\  .
\end{split}
\eeq
We will be interested in keeping terms till $ O(\w)$ which mean we should retain leading/sub-leading terms in $ \mathfrak{G}[d^4]$ and leading term in $\mathfrak{G}[d^3a]$ and
ignore the rest. 

We will now argue that the sub-leading $O(\w)$ terms in $ \mathfrak{G}[d^4]$ do not contribute to the influence phase. The argument is as follows : 
given that $\mathfrak{G}[d^4]$ multiplies a term permutation symmetric in $\w$'s, i.e., $\qd^i(\w_1) \qa^i(\w_2)  \qd^j(\w_3) \qd^j(\w_4)$, its contribution to the 
influence phase should also be permutation symmetric  in $\w$'s. But, the only permutation symmetric linear term in $\w$ is $\w_1+\w_2+\w_3+\w_4$ which vanishes
due to the frequency conserving delta function. Thus, it suffices to examine the leading $O(\w^0)$ term in  $ \mathfrak{G}[d^4]$.

The leading order form is  given by 

\beq
\begin{split}
\mathfrak{G}[d^4] \approx -\frac{\lambda}{2d^3\beta^5}\ 2\pi \ \delta(\sum\w) \ \ointctrclockwise  \frac{d\rho}{2\pi i} \left(2\pi i\frac{d\xi}{d\rho}\right)^2  
\ .
\end{split}
\eeq
Since the integrand is analytic, we can perform this integral by picking up the residue from the  double pole of $\left(\frac{d\xi}{d\rho}\right)^2$. We 
use the Laurent's expansion 
\beq
\left( 2 \pi i \fr{d \xi}{d \rho} \right)^2 \quad = d^2\fr{ \rho^{2d-8}}{(\rho^d-1)^2} \quad =\fr{1}{(\rho-1)^2} - \fr{7-d}{(\rho -1)} + O((\rho-1)^0) \ ,
\eeq
so that we can write 
\beq
\begin{split}
\mathfrak{G}[d^4] \approx (7-d)\frac{\lambda}{2d^3\beta^5}\ 2\pi \ \delta(\sum\w) \ . 
\end{split}
\eeq
We can similarly compute the leading order form of $ \mathfrak{G}[d^3a]$ to get 
\beq
\begin{split}
 \mathfrak{G}[d^3a]\approx 2\be\w \mathfrak{G}[d^4] = 2\be\w(7-d)\frac{\lambda}{2d^3\beta^5}\ 2\pi \ \delta(\sum\w) \ . 
\end{split}
\eeq
 Thus, the derivative expansion for the quartic influence phase
is 
\begin{equation}
\begin{split}
S_I^{(4)}=i^3(7-d)\fr{\lambda}{ 2 d^3 \be ^5}  \int\fr{d^4\w}{(2 \pi)^4} 2\pi \dl(\sum \w)\  \qd^i(\w_1)\qd^i(\w_2)\qd^j(\w_3)\ [\qd^j(\w_4)+2\be \w_4 \qa^j(\w_4)]+\ldots 
\end{split}
\end{equation}
This leading contribution comes entirely from the near horizon region (from the horizon pole), is finite and without any IR divergences. Only two quartic couplings survive in the small \( \w\) limit and  can be identified as the thermal jitter in the damping constant \(\zeta_\gamma \) and the non-Gaussianity of the thermal noise \(\zeta_N\) of \cite{Chakrabarty:2019qcp}. 
There, microscopic time-reversal and KMS conditions relate the corresponding terms in the action up to a factor of \( 2  \be \). Interestingly, our holographic computation matches with this weak coupling result. 

The derivative expansion in the quadratic influence phase is similar. In fact, it is made simpler by the fact that quadratic influence phase can be written as a boundary term (see eq.\eqref{onshell quadratic}).
The only subtlety is that before we take $\rho_c\to \infty$, a temperature-independent  counter-term is needed to get a finite result. We take the counter-term action to be 
\beq
\begin{split}
S^{(2)}_{ct} &=  \left(m_p(T=0) - \fr{r_h \rho_c}{2 \pi \alpha'}\right) \int \fr{d\w}{(2 \pi)} \: \w^2 \qd^i(-\w)\qa^i(\w)\ .
\end{split}
\eeq
We recognise the resulting  influence phase to be that of a particle described by a   non-linear Langevin effective action
\beq
\begin{split}
S_I^{(2)}+ S^{(2)}_{ct} &= m_p \int \fr{d\w}{(2 \pi)} \: \left( \w^2+ i \gamma \w\right) \qd^i(-\w)\qa^i(\w) \\
&\quad+\fr{i}{2} m_p^2 \int \fr{d\w}{(2 \pi)} \:  \left(\langle f^2 \rangle -  \w^2   Z_I  \right)    \qd^i(\w)\qd^i(-\w)+\ldots ,\\
S_I^{(4)} &=i^3  \sum_{ij}\int\fr{d^4\w}{(2 \pi)^4} 2\pi \dl(\sum \w)\  \qd^i(\w_1)\qd^i(\w_2)\qd^j(\w_3)\\ 
&\quad\left[m_p^3\zeta_{\gamma}  \w_4 \qa^j(\w_4)-\fr{1}{4!}m_p^4\zeta_{N}\qd^j(\w_4)\right]+\ldots \ ,
\end{split}
\eeq
where we have the effective couplings of the heavy quark to be
\begin{equation}\label{eq:LangevinCouplings}
\begin{split}
m_p &= m_p(T=0)-\frac{\lambda}{8\pi d\beta }\left(1+ \int_1^\infty\frac{t^{d-4}-1}{t^d-1} dt \right)\ ,\\
m_p^2Z_I &=-\frac{\lambda}{2d^2\beta} \left[\frac{1}{6}+\left(\frac{d}{2\pi}\right)^2
\int_1^\infty \frac{t^{d-4}dt}{t^d-1}\int_1^t \frac{y^{d-4}dy}{y^d-1}(y^4-1)\right]\ ,\\
m_p\gamma &= \frac{\lambda}{2d^2\beta^2} \ ,\quad m_p^2\langle f^2\rangle =  \frac{\lambda}{d^2\beta^3}  \ , \quad
m_p^3\zeta_{\gamma} =(7-d)\frac{ \lambda}{d^3\beta^4}\ ,\quad m_p^4\zeta_{N}=- 12 (7-d) \frac{\lambda }{d^3\beta^5}\ .
\end{split}
\end{equation}
We remind the reader that $\be$ is the inverse Hawking temperature of the black brane. The parameter $\lambda$ is an effective CFT `t Hooft coupling related to the string tension via 
\beq
\begin{split}
\lambda\equiv \frac{16\pi}{\alpha'}\ .
\end{split}
\eeq 
We notice that these results are consistent with the fluctuation-dissipation relations \cite{PhysRev.32.97,PhysRev.32.110,PhysRev.83.34,stratonovich2012nonlinear,Kubo:1957mj,Kubo_1966} which connect $\gamma$ and $\langle f^2\rangle$ on one hand and $\zeta_{\gamma}$ and
$\zeta_{N}$ on the other hand :
\begin{equation}
\begin{split}
\gamma =\frac{1}{2}\beta m_p \langle f^2\rangle\ ,\quad \zeta_{\gamma}= -\frac{1}{12}\beta m_p \zeta_{N}\ .
\end{split}
\end{equation}

We tabulate below the value of the mass corrections and $Z_I$ in various dimensions :
\beq
\begin{array}{||c||c||c||}
\hline
d &   \dl m_p(T)\times \fr{8 \pi d \be}{ \lambda} & Z_I \times \fr{2 d^2 \be}{\lambda}\\
\hline \hline
2 &  0  &   -0.217  \\
\hline \hline 
3 &  \fr{1}{18}(9 \ln 3 - \sqrt{3} \pi) -1 \approx -0.753   &  -0.234 \\
\hline \hline
4 & -1   &     -0.256  \\
\hline \hline
5 & \approx -1.115     &    -0.282       \\
\hline \hline
6 &  \fr{1}{12}(3 \ln 3 - \sqrt{3} \pi) -1 \approx -1.179    &     -0.314    \\
\hline
\end{array}
\eeq

Notice that the thermal mass correction vanishes in \(d=2 \). Our results for \(d=4 \) match with \cite{Son:2009vu,Herzog:2006gh}.

\section{Discussion and Conclusion}\label{sec:discussion}
In this work, we have studied the leading non-linear corrections to the Brownian motion of a heavy quark probing a strongly coupled CFT plasma. This is done using the 
holographic avatar of the Schwinger-Keldysh path integral applied to the world-sheet theory. Our computation relies on a doubled black brane target space and a doubled
string which probes it. This string stretches from the two AdS boundaries and loops around a region obtained by radial Wick rotation that  connects the two stretched horizons.
Evaluating the Nambu-Goto action on this string configuration results in the influence phase of the heavy quark. 

We recognise this influence phase to be that of a particle
with position $q^i$, obeying a non-linear Langevin equation of the form
\begin{equation}
\begin{split}
\frac{d^2q^i}{dt^2}+\Big(\gamma+\zeta_{\gamma}\mathcal{N}^2\Big) \frac{dq^i}{dt} = \langle f^2\rangle \mathcal{N}^i \ .
\end{split}
\end{equation}
Here $\mathcal{N}^2\equiv \mathcal{N}^i \mathcal{N}^i$ and $\mathcal{N}^i$ denotes the thermal noise which is drawn from a non-Gaussian distribution
\begin{equation}
P[\mathcal{N}]\ \propto\ \exp\Big[-\int dt\ \Big(\frac{\langle f^2\rangle}{2}\mathcal{N}^2+\frac{Z_I}{2}\Big(\frac{d\mathcal{N}}{dt}\Big)^2+\frac{\zeta_N}{4!}(\mathcal{N}^2)^2\Big)\Big]\ .
\end{equation}
The parameters can be derived from holography and are given by Eqn.\eqref{eq:LangevinCouplings}.

We notice that these results are consistent with the fluctuation-dissipation relations \cite{PhysRev.32.97,PhysRev.32.110,PhysRev.83.34,stratonovich2012nonlinear,Kubo:1957mj,Kubo_1966} which connect $\gamma$ and $\langle f^2\rangle$ on one hand, and $\zeta_{\gamma}$ and $\zeta_{N}$ on the other hand:
\begin{equation}
\begin{split}
\gamma =\frac{1}{2}\beta m_p \langle f^2\rangle\ ,\quad \zeta_{\gamma}= -\frac{1}{12}\beta m_p \zeta_{N}\ .
\end{split}
\end{equation}
These two relations were derived earlier by studying weakly coupled baths \cite{Chakrabarty:2019qcp}. It is satisfying to see that these relations
continue to hold in holography. We remind the reader that both $\zeta_{\gamma}$ and $\zeta_{N}$ contributions to the influence phase of the heavy quark arise from the correct treatment
of the near horizon region using the prescription of \cite{Glorioso:2018mmw}. The fact that we get a IR finite contribution which is consistent with the non-linear FDT, is a nontrivial check of 
our construction.

 It would be interesting to extend the analysis presented in this paper to account for arbitrary initial states. This would require a detailed analysis of the space of solutions on the holographic SK contour including the  initial time slice data. Such an analysis would extend the existing analyses at zero temperature \cite{Skenderis:2008dg, vanRees:2009rw, Botta-Cantcheff:2015sav, Christodoulou:2016nej, Botta-Cantcheff:2017qir} to finite temperature setup. This would also clarify how different solutions in the horizon cap region may correspond to different initial conditions on the string.

 \par
A possible extension of our work would be to study the effects of backreaction of the string on the black brane geometry. For this, one has to solve Einstein's equations with the appropriate energy-momentum tensor of the string. This corresponds in the CFT to the energy disturbances in the plasma created by the moving quark \cite{Friess:2006fk, gubser2008master, gubser2007energy, gubser2008shock, chesler2008stress, chesler2007wake, gubser2009linearized}. 
 
A different set of subleading effects due to the backreaction are the long time tails. Long time tails are memory effects due to hydrodynamic loops in the boundary plasma \cite{Pomeau:1974hg, Kovtun:2003vj,  CaronHuot:2009iq}. It will be interesting to study such hydrodynamic tails in the CFT and the resulting memory effects in the quark's dynamics as a follow up to our calculations.
 \footnote{We thank participants of Amsterdam String  workshop 2019 for useful discussions on this point.}  
 
 Another possible extension would be to construct a string configuration which will allow us to extract the out-of-time-order influence phase of the heavy quark.
This would require us to generalise our non-linear sigma model to a target space with many copies of the black brane smoothly connected to each other. The studies on weakly
coupled quartic damped oscillator make definite predictions on the relation between the coefficients in the OTO effective theory vs those of the non-linear Langevin theory \cite{Chakrabarty:2019qcp}.
These relations arise from both microscopic time reversal invariance (generalised Onsager relations) and the thermality of the bath (KMS relations). It would be interesting to 
check these relations against a holographic computation. Getting the full OTO influence phase would also be a way to re-derive the result that the quark inherits 
the maximal Lyapunov exponent  from the black brane \cite{deBoer:2017xdk}.

The system studied in this paper permits many generalisations : one could study fermionic strings and the emergence of Fermi-Dirac distributions within holographic Brownian motion.
Or one could  look at strings in rotating blackhole backgrounds. The computation we do in this work can be generalised in different directions : one might ask whether one can replace Brownian strings with Brownian branes and derive their influence phase. One can imagine a sequence of Brownian branes probing the CFT plasma and their effective theories, with the space-filling Brownian brane essentially yielding the
 effective theory of the full plasma. This would give a way to check various features of such effective theories proposed in the literature \cite{Haehl:2015foa}.

  \par
 Another direction to pursue would be to study 
 field theories in the black brane backgrounds and derive dual open quantum field theories for the corresponding single trace primaries, by integrating out the effect of the black brane. All these require extending holographic SK formalism to novel systems and would teach us more about the
structure of real time correlations that emerge out of AdS/CFT.  

The structure of the influence phase we derive in this work  raises many deep conceptual questions that we do not have immediate answers to. We observe that a  
time reversal symmetry breaking order-parameter $\chi$ with a branch cut in its radial dependence plays a crucial  role in our story. The CFT interpretation 
of this object is still unclear to us. Should we think about this as a series of UV modes in the bath which extend till thermal scale that contribute to the influence phase
of the heavy quark ? Perhaps the branch cut is a large N remnant of a line of poles in CFT/bath correlators ? Another question is to trace where the entropy of the 
particle is created from ? Is there an entropy inflow from the complex spacetime that replaces the horizon in holographic SK contour \cite{Haehl:2018lcu,Jensen:2018hhx,Haehl:2018uqv}? We see in this system the glimmer
of many ideas which have recently appeared in the context of fluid SK effective theories \cite{Haehl:2015foa}.

\begin{acknowledgments}
We would like to thank Pinaki Banerjee, Sayantani Bhattacharyya,  Subhobrata Chatterjee, Sudip Ghosh, Akash Jain, Manas Kulkarni, Anupam Kundu, Arnab Kundu, Abhishek Kumar Mehta,  Mukund Rangamani, (Ricardo) Espindola Romero, Amit Sever, Pushkal Shrivastava and Amitabh Virmani for a variety of discussions pertaining to this work. We would also like to acknowledge our debt to the people of India for their steady and generous
support to research in the basic sciences.

\end{acknowledgments}

\appendix

\section{Derivative expansion for Green's function}\label{app:DerivExp}
\subsection{Retarded Bulk to boundary Green's function}
In this subsection, our goal is to derive the structure of the retarded bulk to boundary Green's function. We begin with the following statement : the retarded bulk to boundary Green's function $g_\w$ is the solution of the differential equation
\bea
\frac{d^2g_\w}{d\xi^2}+\be \w \rho \frac{d}{d\xi}[\rho g_\w] =0  
\eea
with the boundary conditions being regularity at the horizon and  $\lim_{\rho\to\rho_c} g_\w =1$. The regularity at the horizon implies 
$\lim_{\rho\to 1}  \frac{dg_\w}{d\xi}=0$.

Setting $g_\w\equiv  e^{\Psi_\w}$,  the differential equation we want to solve takes the form
\bea
\frac{d^2\Psi_\w}{d\xi^2}+\be \w \rho^2 \frac{d \Psi_\w}{d\xi} +\left(\frac{d \Psi_\w}{d\xi}\right)^2 = -\be \w \rho \frac{d\rho}{d\xi}   \ .
\eea
We require the retarded solution of the above equation. The solution can be thought of as a generalisation of Heun function, but its structure is better understood
by working with a small $\omega$ expansion. 

In the static limit, i.e., as $\w \to 0$, we expect $g_\w\to 1$ and $\Psi_\w\to 0$. We can then take 
\bea
\Psi_\w \equiv \ln g_\w \equiv \sum_{k=1}^\infty (-)^k \left(\frac{\be \w}{2}\right)^k  \psi_k \ 
\eea
and substitute this derivative expansion into the above differential equation. This yields the differential equation obeyed by the sequence of functions $\{\psi_k\}$ to be
\beq
\begin{split}
  \frac{d^2\psi_k}{d\xi^2}  =2 \rho^2 \frac{d\psi_{k-1}}{d\xi}-\sum_{m=1}^{k-1}\left(\frac{d\psi_m}{d\xi}\right)\left(\frac{d\psi_{k-m}}{d\xi}\right) \ .
\end{split}
\eeq
This equation also holds for $k=1$ provided we take the convention that $\psi_0 = \ln \rho$.  

The functions $\psi_k$ are those solutions of the above differential equations which decay to zero far away as $\rho\to \rho_c$ and  are regular functions at the horizon (i.e., they are analytic functions without any poles or branch cuts at the horizon). Analyticity in ingoing Eddington Finkelstein co-ordinates is equivalent to setting ingoing boundary conditions at the horizon, as is usually done for computing retarded solutions/correlators. This analycity along with decay at infinity translates to the boundary conditions
\beq
\begin{split}
  \lim_{\rho\to\rho_c} \psi_k  = \lim_{\rho\to 1}  \frac{d\psi_k}{d\xi}=0 \ .
\end{split}
\eeq
 One can then write down an explicit integral which constructs  the solution satisfying these boundary conditions: 
\bea
\psi_k &=& \left(\fr{d}{2 \pi i} \right)^2 \int_{\rho_c}^{\rho} \fr{y^{d-4} dy}{y^d -1}\int_{1}^{y}  \fr{t^{d-4}dt}{t^d -1}\left[2 \rho^2 \frac{d\psi_{k-1}}{d\xi}-\sum_{m=1}^{k-1}\left(\frac{d\psi_m}{d\xi}\right)\left(\frac{d\psi_{k-m}}{d\xi}\right)\right]_{\rho\to t}\ 
\eea
for $k>1$. For $k=1$, we take 
\bea
\psi_1  &=& \left(\fr{d}{2 \pi i}\right)\int_{\rho_c}^{\rho}     \fr{y^{d-4}dy}{y^d -1}   (y^2-1)\ .
\eea
This is the retarded/ingoing solution to the differential equation 
\beq
\begin{split}
 \frac{d^2\psi_1}{d\xi^2} & =2\rho^2 \frac{d\psi_0}{d\xi} = 2\rho \frac{d\rho}{d\xi} \ .
\end{split}
\eeq 
Given such a $\psi_1$, the above integral representation then recursively defines the higher $\psi_k$ functions in terms of lower $\psi_k$ functions.

Note that provided $\lim_{\rho\to 1}  \frac{d\psi_k}{d\xi}=0$, the integrands have no horizon poles and are hence analytic/ingoing at the future horizon. 
Let us look at how the next few orders in derivative expansion works explicitly. We have  
\beq
\begin{split}
\frac{d^2\psi_2}{d\xi^2} &=2\rho^2 \frac{d\psi_1}{d\xi}-\left(\frac{d\psi_1}{d\xi}\right)^2  =\rho^4-1 , \\
 \frac{d^2\psi_3}{d\xi^2} &=2\rho^2 \frac{d\psi_2}{d\xi} -2\left(\frac{d\psi_1}{d\xi}\right)\left(\frac{d\psi_2}{d\xi}\right)=2 \frac{d\psi_2}{d\xi}\ , \\
  \frac{d^2\psi_4}{d\xi^2}&=2\rho^2 \frac{d\psi_3}{d\xi} -2\left(\frac{d\psi_1}{d\xi}\right)\left(\frac{d\psi_3}{d\xi}\right) -\left(\frac{d\psi_2}{d\xi}\right)^2 =2 \frac{d\psi_3}{d\xi}-\left(\frac{d\psi_2}{d\xi}\right)^2\ , \\
  \frac{d^2\psi_5}{d\xi^2} &= 2\rho^2 \frac{d\psi_4}{d\xi} -2\left(\frac{d\psi_1}{d\xi}\right)\left(\frac{d\psi_4}{d\xi}\right)-2\left(\frac{d\psi_2}{d\xi}\right)\left(\frac{d\psi_3}{d\xi}\right) =2 \frac{d\psi_4}{d\xi}-2\left(\frac{d\psi_2}{d\xi}\right)\left(\frac{d\psi_3}{d\xi}\right) \ .
\end{split}
\eeq
The corresponding integrals are then given by 
\bea
\psi_2 &=& \left(\fr{d}{2 \pi i} \right)^2 \int_{\rho_c}^{\rho} \fr{y^{d-4} dy}{y^d -1}\int_{1}^{y}  \fr{t^{d-4}dt}{t^d -1}   (t^4-1)\ ,  \\
\psi_3 &=& \left(\fr{d}{2 \pi i} \right)^2 \int_{\rho_c}^{\rho} \fr{y^{d-4} dy}{y^d -1}\int_{1}^{y}  \fr{t^{d-4}dt}{t^d -1}   \left[2 \frac{d\psi_2}{d\xi}\right]_t\ ,\\
\psi_4 &=& \left(\fr{d}{2 \pi i} \right)^2 \int_{\rho_c}^{\rho} \fr{y^{d-4} dy}{y^d -1}\int_{1}^{y}  \fr{t^{d-4}dt}{t^d -1} \left[2 \frac{d\psi_3}{d\xi}-\left(\frac{d\psi_2}{d\xi}\right)^2\right]_t\ ,  \\
\psi_5 &=& \left(\fr{d}{2 \pi i} \right)^2 \int_{\rho_c}^{\rho} \fr{y^{d-4} dy}{y^d -1}\int_{1}^{y}  \fr{t^{d-4}dt}{t^d -1}\left[2  \frac{d\psi_4}{d\xi}-2\left(\frac{d\psi_2}{d\xi}\right)\left(\frac{d\psi_3}{d\xi}\right)\right]_t\ .
\eea

Reality condition then implies 
\beq
\psi_k^\ast = (-)^k\psi_k\ .
\eeq

\subsection{Derivative expansion of the linearised solution}
One can expand the exponentials (including the Bose Einstein factors ) and write down a derivative expansion for the linearised solution $\X^i$ in the form 
\beq
\begin{split}
\X^i &\equiv \X^i_{(0)} + \X^i_{(1)} + \X^i_{(2)} + \X^i_{(3)} + \X^i_{(4)} +\ldots\ ,
\end{split}
\eeq
where 
\beq
\begin{split}
\X^i_{(0)} &\equiv \xi\ \qR^i  + (1-\xi)\ \qL^i\ , \\
\X^i_{(1)} &\equiv \fr{\be \w}{2}\left\{ \xi(1-\xi)(\qR^i-\qL^i) - \psi_1 \X^i_{(0)}  \right\}\ ,\\
\X^i_{(2)} &\equiv \left( \frac{1}{2 !} \psi_1^2 +  \psi_2 \right)\left(\fr{\be \w}{2} \right)^2\X^i_{(0)}-\left(\fr{\be \w}{2}\right)^2 (\qR^i-\qL^i)\bigg\{ \psi_3 \\ 
  &\ + \psi_1 \xi(1-\xi) + \frac{1}{3} \xi(1-\xi)(2\xi-1) \bigg\}\ ,\\
\X^i_{(3)} &\equiv -\left( \frac{1}{3 !} \psi_1^3 +  \psi_1 \psi_2 + \psi_3 \right)\left(\fr{\be \w}{2}  \right)^3\X^i_{(0)}+\left(\fr{\be \w}{2}\right)^3 (\qR^i-\qL^i)\bigg\{ \psi_1\psi_3 + \psi_3(2\xi-1)\\
  &\ +\left( \frac{1}{2 !} \psi_1^2 +  \psi_2 \right) \xi(1-\xi) +  \frac{\psi_1}{3} \xi(1-\xi)(2\xi-1)-\frac{1}{3} \xi^2(1-\xi)^2 \bigg\}\ ,\\
\X^i_{(4)} &\equiv \left( \frac{1}{4 !} \psi_1^4 +  \frac{1}{2 !} \psi_1^2 \psi_2 +  \frac{1}{2 !} \psi_2^2 + \psi_1 \psi_3 +\psi_4 \right)\left( \fr{\be \w}{2}\right)^4\X^i_{(0)}-\left(\fr{\be \w}{2}\right)^4 (\qR^i-\qL^i)\bigg\{\psi_5  \\
   &\ + \psi_2 \psi_3 +   \frac{1}{2 !} \psi_1^2 \psi_3+\frac{1}{3} \psi_3 + \psi_1 \psi_3(2\xi-1) + \left( \frac{1}{3 !} \psi_1^3 +\psi_1  \psi_2 -\psi_3\right) \xi(1-\xi)  \\
  &\  +  \frac{1}{3} \xi(1-\xi)(2\xi-1) \left(\frac{1}{2} \psi_1^2+\psi_2-\frac{1}{15}\right) -\frac{1}{3} \xi^2(1-\xi)^2 \psi_1 -\frac{1}{15} \xi^2(1-\xi)^2 (2\xi-1)  \bigg\} \ .
\end{split}
\eeq

\section{Exact solution in AdS\({}_3\)}\label{app:AdS3}

In the case of AdS$_3$, the retarded Green's function takes the form
\bea
g_\w=\left(\fr{\rho_c}{\rho}\right)\fr{\w+i\ r_h \rho}{\w+i\ r_h \rho_c}\ ,\qquad \chi\equiv \xi+\psi_1=\frac{1}{2\pi i} \ln\left[\fr{\rho-1}{\rho+1}\times\fr{\rho_c+1}{\rho_c-1}\right]\ . 
\eea

\bea
e^{\be\w \chi}=\left[\fr{\rho-1}{\rho+1}\times\fr{\rho_c+1}{\rho_c-1}\right]^{\frac{\be\w}{2\pi i}}\ . 
\eea
Expanding this $g_\w$ in the derivative expansion, we obtain
\bea
\psi_k=\fr{1}{k}\left(\fr{d}{2\pi i}\right)^k \left( \fr{1}{\rho_c^k}-\fr{1}{\rho^k}\right)\ .  
\eea
This can be used to get a closed form expression for $\xi$ as
\bea
\xi=\frac{1}{2\pi i} \ln\left[\fr{\rho-1}{\rho+1}\times\fr{\rho_c+1}{\rho_c-1}\right]-\fr{1}{\pi i} \left( \fr{1}{\rho_c}-\fr{1}{\rho}\right)\ . 
\eea

We can also compute the influence phase in this case. We have 
\beq
\begin{split}
A_\omega= 
-2\pi i \frac{\w}{r_h}\rho_c\ \frac{\rho\w+\frac{i}{2}r_h(1+\rho^2)}{\w+i\ r_h \rho_c}\ , \qquad
B_\omega= 
-\pi \w\rho_c\ \frac{\rho^2-1}{\w+i\ r_h \rho_c}\ . 
\end{split}
\eeq
and
\beq
\begin{split}
e^{-\be\w \chi} &=\left[\fr{\rho-1}{\rho+1}\times\fr{\rho_c+1}{\rho_c-1}\right]^{-\frac{\be\w}{2\pi i}}\ , \qquad
 \frac{d\xi}{d\rho}= \frac{1}{i\pi} \frac{1}{\rho^2 (\rho^2-1)}\ .
 \end{split}
\eeq

As an example, we can compute the exact quartic influence phase  in AdS$_3$.
The quartic influence phase  of a quark moving through strongly interacting CFT$_2$ plasma can be computed explicitly to be 
\beq
\begin{split}
 \mathfrak{G}[PFFF] \equiv &\  2\frac{(2\pi\rho_c)^4\lambda}{2^7\beta^5}\left(\fr{\w_1}{r_h\rho_c+i\w_1}\right)
\left(\fr{\w_2}{r_h\rho_c-i\w_2}\right) 
\left(\fr{\w_3}{r_h\rho_c-i\w_3} \right)
\left(\fr{\w_4}{r_h\rho_c-i\w_4}\right)\\
&\   2\pi \ \delta(\sum\w) \ \ointctrclockwise  \frac{d\rho}{2\pi i} \left(2\pi i\frac{d\xi}{d\rho}\right)^2\  \mathcal{I}_{PFFF}\ e^{\be\w_1(1-\chi)} \ ,\\
\mathfrak{G}[PPPF] \equiv &\  2\frac{(2\pi\rho_c)^4\lambda}{2^7\beta^5}\left(\fr{\w_1}{r_h\rho_c+i\w_1}\right)
\left(\fr{\w_2}{r_h\rho_c+i\w_2}\right) 
\left(\fr{\w_3}{r_h\rho_c+i\w_3} \right)
\left(\fr{\w_4}{r_h\rho_c-i\w_4}\right)\\
& \  2\pi \ \delta(\sum\w) \ointctrclockwise  \frac{d\rho}{2\pi i} \left(2\pi i\frac{d\xi}{d\rho}\right)^2  \mathcal{I}_{PPPF}\ e^{-\beta\w_4 (1-\chi)} \ ,\\
\mathfrak{G}[PFPF] \equiv &\  -2\frac{(2\pi\rho_c)^4\lambda}{2^7\beta^5}\left(\fr{\w_1}{r_h\rho_c+i\w_1}\right)
\left(\fr{\w_2}{r_h\rho_c-i\w_2}\right) 
\left(\fr{\w_3}{r_h\rho_c+i\w_3} \right)
\left(\fr{\w_4}{r_h\rho_c-i\w_4}\right) \\
&\  2\pi \ \delta(\sum\w) \ointctrclockwise  \frac{d\rho}{2\pi i} \left(2\pi i\frac{d\xi}{d\rho}\right)^2\   \mathcal{I}_{PFPF}\ e^{\be(\w_1+\w_3)(1-\chi)} \ ,\\
\mathfrak{G}[PPFF] \equiv &\  -\frac{(2\pi\rho_c)^4\lambda}{2^7\beta^5}\left(\fr{\w_1}{r_h\rho_c+i\w_1}\right)
\left(\fr{\w_2}{r_h\rho_c+i\w_2}\right) 
\left(\fr{\w_3}{r_h\rho_c-i\w_3} \right)
\left(\fr{\w_4}{r_h\rho_c-i\w_4}\right)\\
&\  2\pi \ \delta(\sum\w)  \ointctrclockwise \frac{d\rho }{2\pi i}  \left(2\pi i\frac{d\xi}{d\rho}\right)^2  \mathcal{I}_{PPFF}\ e^{\be(\w_1+\w_2)(1-\chi)} \ .
\end{split}
\eeq
Here we have 
\beq
 \left(2\pi i\frac{d\xi}{d\rho}\right)^2 =\fr{4}{\rho^4(\rho^2-1)^2}\ ,\quad
e^{\be\w \chi}=\left[\fr{\rho-1}{\rho+1}\times\fr{\rho_c+1}{\rho_c-1}\right]^{\frac{\be\w}{2\pi i}}\ .
\eeq
The functions that appear in this influence phase are given by the expressions
\beq
\begin{split}
\mathcal{I}_{PFFF} &\equiv \rho^8-1 +
i\ (\w_1-\w_3-\w_4) \left(\fr{\be\rho}{2\pi}\right)(\rho^4-1)(\rho^2+1)-i\ \w_2 \left(\fr{\be\rho}{2\pi}\right)(\rho^2-1)^3\\
&+2 \left(\fr{\be\rho}{2\pi}\right)^2 (\w_1\w_3+\w_1\w_4-\w_3\w_4)(\rho^4-1)
- 4i \left(\fr{\be\rho}{2\pi}\right)^3(\rho^2-1)\w_1 \w_3 \w_4\ ,\\
\mathcal{I}_{PPPF} &\equiv \rho^8-1 +
i\ (\w_1-\w_3-\w_4) \left(\fr{\be\rho}{2\pi}\right)(\rho^4-1)(\rho^2+1)
+2i\ \w_3 \left(\fr{\be\rho}{2\pi}\right)(\rho^4+1)(\rho^2-1)\\
&+2 \left(\fr{\be\rho}{2\pi}\right)^2 (\w_1\w_4+\w_2\w_4-\w_1\w_2)(\rho^4-1)
+ 4i \left(\fr{\be\rho}{2\pi}\right)^3(\rho^2-1)\ \w_1 \w_2 \w_4\ ,\\
\mathcal{I}_{PFPF} &\equiv (\rho^4-1)^2 +
i\ (\w_1-\w_2+\w_3-\w_4) \left(\fr{\be\rho}{2\pi}\right)(\rho^2-1)^2(\rho^2+1) \\
&+2 \left(\fr{\be\rho}{2\pi}\right)^2 (\w_2\w_3+\w_1\w_4)(\rho^2-1)^2\ ,\\
\mathcal{I}_{PPFF} &\equiv \rho^8+6\rho^4+1 +i\ (\w_1+\w_2-\w_3-\w_4) \left(\fr{\be\rho}{2\pi}\right)(\rho^2+1)^3\\
&+2 \left(\fr{\be\rho}{2\pi}\right)^2 (\rho^2+1)^2 [(\w_1+\w_2)(\w_3+\w_4)-\w_1\w_2-\w_3\w_4]\\
&+ 4i \left(\fr{\be\rho}{2\pi}\right)^3(\rho^2+1)\ [\w_1 \w_2 (\w_3+\w_4)-\w_3\w_4(\w_1+\w_2)]
+ 8 \left(\fr{\be\rho}{2\pi}\right)^4\ \w_1 \w_2 \w_3\w_4\ .
\end{split}
\eeq

\bibliographystyle{JHEP}
\bibliography{BrownianAdSBib}

\end{document}